\documentclass[conference]{IEEEtran}
\IEEEoverridecommandlockouts
\usepackage{cite}
\usepackage{amsmath,amssymb,amsfonts}
\usepackage{algorithmic}
\usepackage{graphicx}
\usepackage{textcomp}
\usepackage{xcolor}
\usepackage{mathtools,dsfont}
\usepackage{psfrag,bbm,graphicx}
\usepackage{comment,balance}
\usepackage{tikz}
\usepackage{url}

\usepackage{epsfig,subfig}
\usepackage{multicol}
\usepackage{multirow}
\usepackage{amsfonts, color}
\usepackage{array}
\usepackage{amssymb}
\usepackage[linesnumbered,ruled,vlined]{algorithm2e}
\usepackage{nccmath}
\usepackage[belowskip=-10pt,aboveskip=0pt]{caption}
\newtheorem{lemma}{Lemma}
\newtheorem{remark}{Remark}

\newtheorem{theorem}{Theorem}

\newtheorem{definition}{Definition}

\newtheorem{assumption}{Assumption}

\begin{document}

\title{On Renting Edge Resources for Service Hosting
}
\author{V S Ch Lakshmi Narayana, Sharayu Moharir, and Nikhil Karamchandani \\  Department of Electrical Engineering, IIT Bombay}
\maketitle
\begin{abstract}
The rapid proliferation of shared edge computing platforms has enabled application service providers to deploy a wide variety of services with stringent latency and high bandwidth requirements. A key advantage of these platforms is that they provide pay-as-you-go flexibility by charging clients in proportion to their resource usage through short-term contracts. This affords the client significant cost-saving opportunities, by dynamically deciding when to host its service on the platform, depending on the changing intensity of requests. 

A natural policy for our setting is the Time-To-Live (TTL) policy. We show that TTL performs poorly both in the  adversarial arrival setting, i.e., in terms of the competitive ratio, and for i.i.d. stochastic arrivals with low arrival rates,  irrespective of the value of the TTL timer.  

We propose an online policy called RetroRenting (RR) and show that in the class of deterministic online policies, RR is order-optimal with respect to the competitive ratio. 

In addition, we provide performance guarantees for RR for i.i.d. stochastic arrival processes  coupled with  negatively associated rent cost sequences and prove that it compares well with the optimal online policy. Further, we conduct simulations using both synthetic and real world traces to compare the performance of RR  with the optimal offline and online policies. The simulations show that the performance of RR is near optimal for all settings considered. Our results illustrate the universality of RR.

\end{abstract}

\maketitle
\section{Introduction}
Widespread adoption of smartphones and other handheld devices over the last decade has been accompanied with the development of a wide variety of mobile applications providing a plethora of services.\footnote{ A preliminary version of this work appeared in WiOpt 2020 \cite{narayana2020retrorenting}. } These applications often rely on cloud computing platforms \cite{fernando2013mobile} to enable the delivery of high-quality performance anytime, anywhere to resource-constrained mobile devices. However, the last few years have seen the emergence of applications based on machine learning, computer vision, augmented/virtual reality (AR/VR) etc. which are pushing the limits of what cloud computing platforms can reliably support in terms of the required latency and bandwidth. This is largely due to the significant distance between the end user and the cloud server, which has led the academia and the industry to propose a new paradigm called edge computing  \cite{satyanarayanan2017emergence} whose basic tenet is to bring storage and computing infrastructure closer to the end users. This can help enable applications with ultra-small network latency and/or very high bandwidth requirements, which cannot be reliably supported by the backhaul connection. As a concrete example, consider a user in a wildlife sanctuary, capturing the scene around her live on a mobile device, which relays the image/video to an edge server. Using its much higher computational and storage capabilities, an application on the edge server can continually detect species of plants, animals, birds and relay this information back to the end user device where it can be overlaid onto the live stream to provide a much richer viewing experience. Broader applications of edge computing include industrial robotics/drone automation, AR/VR-based infotainment and gaming, autonomous driving and the Internet of Things (IoT). While there are now several industry offerings of dedicated edge computing platforms, e.g., Amazon Web Services \cite{aws} and Microsoft Azure \cite{azure}, there have also been proposals to augment cellular base stations \cite{chen2017collaborative} and WiFi access points \cite{willis2014paradrop} so that they can act as edge servers.

As discussed in \cite{jiang2020economic}, a typical system with edge-computation capabilities has the followings stakeholders: \emph{(i)} application providers who provide a service to the end-user, \emph{(ii)} access providers who rent out edge resources to the application providers to host their services to serve their customers, and \emph{(iii)} Internet service providers (ISPs) who provide the backbone network service to the application providers. The focus of this work is on the algorithmic task for an application provider of determining when to rent edge resources and how to use the backbone network to best serve their customers while minimizing the cost of service. Although we study this problem from a perspective of a specific application provider, the effect of the presence of other application providers who might be  simultaneously interested in using the resources offered by the access providers is captured through the time-varying nature of the cost of renting edge resources. More specifically, we consider the setting where the access provider periodically revises the cost of renting edge resources, potentially based on the current demand, and the application provider decides if they wish to rent edge resources at the quoted cost.

In this work, we say that a service is hosted  at an edge server, if all the data and code needed to run the service has been downloaded from a remote/back-end server (possibly in the cloud) and hosted  on the edge server. Thus the edge server can handle service requests on its own without requiring to communicate with the back-end server  using the backbone network.  Edge servers are often limited in computational capability as compared to cloud servers \cite{tran2019costa}, and hence there might be a limit on the number of parallel requests they can serve for the hosted  service. An application provider can avail this ability to host  on the edge server in return of a  potentially time-varying cost  in proportion to the amount of resources used and/or the duration of rental.  Since computing platforms usually provide pay-as-you-go flexibility \cite{mouradian2017comprehensive}, the client can dynamically decide when to host  or evict the service at the edge, depending on the varying number of arriving service requests. The application provider needs to design an efficient service hosting policy which can help minimize the overall cost of deploying the service. 


Each time a service request arrives at the edge server and the service is not hosted  on it, there are two options: \textit{(a) request forwarding} which simply forwards the service request to the back-end server, which then carries out all the relevant computation for addressing the request; and \textit{(b) service download} which downloads all the data and code needed for running the service from the back-end server and hosts  it at the edge. The cost for these two actions is different, depending for example on the amount of network bandwidth needed or the latency incurred for each of them. Motivated by empirical evidence \cite{satyanarayanan2017emergence, ha2017you}, a natural assumption is that the cost of forwarding a single request to the back-end server is lower than the cost of downloading the entire service to host  it on the edge server \cite{zhao2018red}. 

Our goal in this work is to design online service hosting  policies for the application provider which aim to minimize the total cost it incurs for serving requests, which is a combination of the request forwarding cost, the service download cost and the edge server rental cost. We consider two classes of request arrival processes: (i) \textit{adversarial arrivals}: the request sequence is arbitrary and the performance of any online policy is measured by its competitive ratio, which provides a worst-case guarantee on its performance for any request arrival sequence in comparison to the optimal offline policy, which has knowledge of the entire arrival sequence a-priori. (ii) \textit{i.i.d. stochastic arrivals}: requests are generated according to an i.i.d. stochastic process and we compare the expected cost of a proposed policy with that of the optimal online policy  which does not have knowledge of sample path of future arrivals, but can exploit the statistics of the stochastic process to make service hosting decisions. 

\subsection{Our Contributions}

A natural policy for our setup is the Time-To-Live (TTL) policy \cite{carra2019ttl}, which is popular in the content caching literature. Under the TTL policy, each ``miss" triggers a download of the service to the edge server where it is then retained for some fixed amount of time. We show that TTL performs poorly both in terms of the competitive ratio for arbitrary arrivals and for i.i.d. stochastic arrivals with low arrival rates, irrespective of the value of the TTL timer. Given the limitations of TTL, we propose an online policy called RetroRenting (RR) which uses the history of request arrivals  and the history of rent costs  to decide when to host  or evict the service at the edge server. Under the adversarial request setting, we show that RR is order-optimal with respect to the competitive ratio in the class of deterministic online policies. In addition, we also provide performance guarantees for RR under  i.i.d. request arrivals and negatively associated rent costs  and  prove that it compares very well with the optimal online policy in this setting. In addition to our analytical results, we conduct simulations using both synthetic and real world traces to compare the performance of RR  with the optimal offline and online policies. Our simulations show that the performance of RR is near optimal for all settings considered. These results combined illustrate the universality of RR.

\subsection{Related Work} 

Mobile applications have increasingly become more and more demanding in terms of their bandwidth and latency requirements. This, along with the advent of new time-critical applications such as the Internet of Things (IoT), AR/VR and autonomous driving has necessitated the migration of a part of the storage and computing capabilities from remote servers to the edge of the network. See \cite{Puliafito:2019, mao2017survey, mach2017mobile} for a survey of various edge computing architectures and proposed applications.
The emergence of such edge computing platforms \cite{Puliafito:2019, mao2017survey, mach2017mobile} has been accompanied with various academic works which model and analyse the performance of such systems. We briefly discuss some of the relevant works in the literature. 

One approach towards designing efficient edge computing systems is to formulate the design problem as a large one-shot static optimization problem which aims to minimize the cost of operating the edge computing platform  \cite{pasteris2019service, bi2019joint, chen2017collaborative, tran2019costa, yang2015cost}. \cite{pasteris2019service} considers such a problem in a heterogeneous setting where different edge nodes have different storage or computation capabilities and various services have different requirements. The goal is to find the optimal service placement scheme subject to the various constraints. The authors show that the problem is in general NP-hard and  propose constant factor approximation algorithms. A similar problem is considered in \cite{bi2019joint} which looks at the setting where an edge server is assisting a mobile unit in executing a collection of computation tasks. The question of which services to cache at the edge and which computation tasks to offload are formulated as a mixed integer non-linear optimization problem and the authors design a reduced-complexity alternative minimization based iterative algorithm for solving the problem. Similar problems have also been considered in \cite{chen2017collaborative, tran2019costa, yang2015cost}. Our work differs from this line of work in that we are interested in designing online algorithms which adapt their service placement decisions over time depending on the varying number of requests. 

One approach to modeling time-varying requests is to use a stochastic model as done in \cite{xu2018joint, chen2019budget, wang2015dynamic} which assumes that requests follow a Poisson process and then attempts to minimize  the computation latency in the system by optimizing the service hosting and task offloading decisions. \cite{chen2019budget} considers a setting where the underlying distribution for the request process is a-priori unknown and uses the framework of Contextual Combinatorial Multiarmed Bandits to learn the demand patterns over time and make appropriate  decisions. Finally, \cite{wang2015dynamic} considers a Markovian model for user mobility and uses a Markov Decision Process (MDP) framework to decide when and which services to migrate between different edge servers as the users move around. Our work differs from these works in that in addition to stochastic request models, we also focus on the case of arbitrary request arrival processes and provide `worst-case' guarantees on the performance of our proposed schemes instead of `average' performance guarantees. This can be vital in scenarios where the arrival patterns change frequently over time, making it difficult to predict demand or model it well as a stochastic process.

The work closest to ours is \cite{zhao2018red} which considers an edge server with  limited memory $K$ units which can be used at zero cost and an arbitrary request process for a catalogue of services. This work studies the design of service caching policies which minimize the cost incurred by the edge server for deploying the various services. The authors propose an online algorithm called ReD/LeD and prove that the competitive ratio of the proposed scheme is at most $10K$. Unlike \cite{zhao2018red}, we study the problem from the perspective of an application provider and design cost-efficient service hosting  policies which dynamically decide when to cache or evict the service at the edge. In addition to  worst-case performance guarantees, we provide average performance guarantees of our proposed policy.

 Online algorithms have been studied for a wide variety of computational problems \cite{10.5555/290169}. In particular, \cite{10.1145/146585.146588} studies a general compute system for the processing of a sequence of tasks, each of which requires the system to be in a certain state for execution. There is a cost metric governing the penalty for moving the system from one state to the other, and the goal is to design online schemes which have a good competitive ratio with respect to the offline optimal. While the setting in \cite{10.1145/146585.146588} is similar to our work, the assumptions made there on the cost metrics are not satisfied for our problem and hence their results are not applicable.
 
 Other relevant works on exploiting edge resources for service include \cite{xu2020collaborate} in which the focus is on the problem of characterizing the benefits of sharing resources for service caching on the edge server with multiple network providers. In \cite{wei2020mobility}, the focus is on decision pro-active service caching to serve users with high-mobility. Further, \cite{jiang2020economic, zeng2020novel} focus on economic aspects of edge caching involving interactions between different stakeholders using game-theoretic tools. In \cite{zhang2020cooperative}, the authors characterize the benefits of cooperation between edge servers and propose a deep reinforcement based algorithm for effective cooperation. 

Finally, as mentioned before, the problem of service caching does resemble the content caching problem but with some key differences. Content caching has a rich history, see for example \cite{borst2010distributed, tan2012optimal, Wolman99, breslau1999web, sleator1985amortized, belady1966study}. A popular class of online caching policies is the Time-To-Live (TTL) policy \cite{carra2019ttl}, which downloads a content to the cache upon a cache miss and then retains it there for a certain fixed amount of time. In this work, we consider a variant of the TTL policy for service caching and demonstrate that it performs poorly in several cases.

\section{System Setup}
\label{sec:setting}
\subsection{Network Model}\label{syst:nwmodel}
We study a system consisting of a back-end server and an edge
server in proximity to the end-user. The back-end server always
stores the service. The service can be hosted on the edge server by
paying a rental cost. On paying this cost, requests can be served
at edge free of cost, subject to an upper limit on the number of
concurrent requests  of our service  being served at the edge. In addition, requests
can be served by the back-end server at a non-zero cost. The back-end server can serve all the requests that are routed to it.
\subsection{Arrival Process  and rent costs}
We consider a time-slotted system and consider both adversarial and stochastic settings. In the adversarial setting, we make no assumptions on the arrival sequence and rent cost sequence. In the stochastic setting, we make the following assumption. 

\begin{assumption}\label{assum_stochastic}(i.i.d. stochastic arrivals and negatively associated rent costs)
The number of requests arriving in a time-slot is independent and identically distributed across time-slots. More specifically, let $X_t$ be the number of requests arriving in time-slot $t$. Then, for all $t$,
$$\mathbb{P}(X_t = x) = p_x \text{ for } x = 0, 1, 2, \cdots.$$

The amount of rent costs to be paid for storage in each time-slot are random and negatively associated across time-slots \cite{wajc2017negative}, for example i.i.d sequence rent costs across time slots is special case of negative associated random sequence. Let $R_t$ be the rent cost per time-slot, $\{R_t\}_{t\geq 1}$ is the sequence of negatively associated random variables and $c= \mathbb{E}[R_t]$.

\end{assumption}

\subsection{Sequence of Events in a Time-slot}
The following sequence of events occurs in each time-slot. We first have request arrivals. If the service is hosted on the edge server, requests are served locally subject to the constraints on the  allocated computation power of the edge server, else requests are forwarded to the back-end server.  
After that, we know the amount of rent cost in the immediate next time-slot.  The system then makes a hosting decision (fetch/evict/no change).

\subsection{Cost Model and Constraints}\label{syst:costmodel}
Our cost model  builds on the model proposed in \cite{narayana2020retrorenting, zhao2018red} and extends it to the setting where edge resources can be rented in a dynamic manner by paying a potentially time-varying  rent cost. 
For a given policy $\mathcal{P}$, the total cost incurred in time-slot $t$, denoted by $C_t^{\mathcal{P}}$, is the sum of the following three costs.
\begin{enumerate}
	\item[--] \emph{Service cost $(C_{S,t}^\mathcal{P})$}: Each request forwarded to the back-end server is served at the cost of one unit. 
	\item[--] \emph{Fetch cost $(C_{F,t}^\mathcal{P})$}: On each fetch of the service from the back-end server to host on the edge-server, a fetch cost of $M(>1)$ units is incurred. 
	\item[--] \emph{Rent cost $(C_{R,t}^\mathcal{P})$}: A rent cost of $c_t $ units is incurred to host  the service on the edge server for a time-slot. Here,  
	$c_t$ lies in $[c_{\text{min}}, c_{\text{max}}]$ with $c_{\text{min}}>0$.	
\end{enumerate}

\begin{remark}
	We model the rent cost as a time-varying quantity as it might depend on various factors including the overall demand for edge resources across various services that are potentially interested in renting. The time-varying nature of the rent cost is the key difference between this work and the model considered in its preliminary version in \cite{narayana2020retrorenting}.
\end{remark}

Since the edge server offers a limited computation power per service, the number of requests that can be served by the edge server in a time-slot is limited to $\kappa \in \mathbb{Z}^+$, where $\mathbb{Z}^+$ is the set of all positive integers. 
Let $r_t$ be an indicator of the event that the service is hosted on the edge server during time-slot $t$. It follows that
\begin{align}
\label{equation:total_cost}
C_t^\mathcal{P} &=C_{S,t}^\mathcal{P}+C_{F,t}^\mathcal{P}+C_{R,t}^\mathcal{P}, \\
\text{where, } C_{S,t}^\mathcal{P} & =
\begin{cases*}
X_t-\min\{X_t,\kappa\} & \text{ if $r_t = 1$ } \nonumber \\
X_t & \text{ otherwise,} 
\end{cases*} \nonumber\\
C_{F,t}^\mathcal{P}&=
\begin{cases*}
M & \text{ if $r_{t} = 0$ and $r_{t+1} = 1$ } \nonumber\\
0 &\text{ otherwise,} 
\end{cases*}\\
C_{R,t}^\mathcal{P} &=
\begin{cases*}
c_t & \text{ if $r_{t} = 1$ } \nonumber \\
0 & \text{ otherwise.} 
\end{cases*}
\end{align}

\begin{remark}
	We limit our discussion to the case where the rent cost $c_t\leq M+\kappa$ because, for $c_t>M+\kappa$, it is optimal to forward all requests to the back-end server, irrespective of the value of $M$ and the arrival sequence.
\end{remark}

\subsection{Algorithmic Challenge}
The algorithmic challenge is to design a policy which decides when to host  the service on  edge server. 
Policies can be divided into the following two classes. 
\begin{definition}(Types of Policies)
	\label{defn:typesOfPolicies}
	\begin{enumerate}
		\item[--] \emph{Offline Policies}: A policy in this class knows the entire request arrival sequence and rent cost sequence a-priori.
		\item[--] \emph{Online Policies}: A policy in this class does not have knowledge of future arrivals  and the cost of renting across future time-slots. 
	\end{enumerate}
\end{definition}
We design an online policy which makes decisions based on the request arrivals, rent costs thus far and the various costs and constraints, the fetch cost $(M)$, and the edge server constraint $(\kappa)$. 
\subsection{Metric and Goal}
The optimal offline and online policies serve as benchmarks to evaluate the performance of the proposed policy.  
We use different cost metrics for the adversarial and stochastic request arrival settings. 
\subsubsection{Adversarial setting} For the adversarial setting, we compare the performance of a policy $\mathcal{P}$ with the performance of the optimal offline policy (OPT-OFF). The goal is to design a policy $\mathcal{P}$ which minimizes the competitive ratio $\rho^{\mathcal{P}}$ defined as 
\begin{equation}
\label{eq:competitiveRatio}
 \rho^{\mathcal{P}}=\sup_{a\in \mathcal{A}, b\in \mathcal{R}} \frac{C^{\mathcal{P}}(a,b)}{C^{\text{OPT-OFF}}(a,b)},
\end{equation}

where $\mathcal{A}$ is the set of all possible finite request arrival sequences,   $\mathcal{R}$ is the set of all possible finite rent cost sequences, $C^{\mathcal{P}}(a,b)$, $C^{\text{OPT-OFF}}(a,b)$ are the overall costs of service for the request arrival sequence $a$ and rent cost sequence $b$   under online policy $\mathcal{P}$ and the optimal offline policy respectively. 
\subsubsection{i.i.d. stochastic arrivals and  negatively associated rent costs } For i.i.d. stochastic arrivals and  negatively associated rent costs   (Assumption \ref{assum_stochastic}), we compare the performance of a policy $\mathcal{P}$ with the performance of the optimal online policy (OPT-ON).
%
The goal is to minimize $\sigma^{\mathcal{P}}_T$, defined as the ratio of the expected cost incurred by policy $\mathcal{P}$ in $T$ time-slots to that of the optimal online policy in the same time interval. Formally,
\begin{equation}
\label{eq:efficiencyRatio}
\sigma^\mathcal{P}(T) =\frac{\mathbb{E}\bigg[\displaystyle\sum_{t=1}^T C_t^\mathcal{P}\bigg]}{\mathbb{E}\bigg[\displaystyle\sum_{t=1}^T C_t^{\text{OPT-ON}}\bigg]},
\end{equation}
where $C_t^\mathcal{P}$ is as defined in \eqref{equation:total_cost}.
\section{Main Results and Discussion}
In this section, we state and discuss our main results. We provide outlines of the proofs in Section \ref{sec:proofOutlines} and the details of the proofs are discussed in Section \ref{sec:proofs}. 


\subsection{Our Policy: RetroRenting (RR)} \label{subsec:Retro_renting}

A policy determines when to fetch and host  the service and when to evict the service from the edge. The RR policy makes these decisions in each time-slot by evaluating if it made the right decision in hindsight. We first provide an overview of the RR policy. 
	
	\emph{To fetch}: Let the service not be hosted  at the beginning of time-slot $t$ and $t_{\text{evict}}<t$ be the time when the service was most recently evicted by RR. The RR policy searches for a time-slot $\tau$ such that $t_{\text{evict}} < \tau < t$, and the total cost incurred is lower if the service is fetched in time-slot $\tau-1$ and hosted  during time-slots $\tau$ to $t$ than if the service is not hosted  during time-slots $\tau$ to $t$. If there exists such a time $\tau$, the RR policy fetches the service in time-slot $t$.

	\emph{To evict}: Let the service be hosted  at the edge at the beginning of time-slot $t$ and $t_{\text{fetch}}<t$ be the time when the service was most recently fetched by RR. 
	The RR policy searches for a time-slot $\tau$ such that $t_{\text{evict}} < \tau < t$, and the total cost incurred is lower if the service is not hosted  at the edge during time-slots $\tau$ to $t$ and fetched in time-slot $t$ than if the service is hosted  at the edge during time-slots $\tau$ to $t$. If there exists such a time $\tau$, the RR policy evicts the service in time-slot $t$.

Refer to Algorithm \ref{algo:RR} for a formal definition of the RR policy. The notation used in Algorithm \ref{algo:RR} is summarized in Table \ref{table:RR}.
\begin{table}[h]
	\centering
	\begin{tabular}{ |c | l|} 
		\hline
		\textbf{Symbol}&  \textbf{Description} \\ 
		\hline
		\hline
		$t$ & Time index\\
		\hline
		$M$ & Fetch cost\\
		\hline
		$c_t$       & Rent cost in time-slot $t$\\
		\hline
		$c_{\text{min}}$       &Minimum value of $c_t$\\
		\hline
		$c_{\text{max}}$       &Maximum value of $c_t$\\
		\hline
		$\kappa$  & Maximum number of  our service  requests  that \\
		& can be served by the edge server in a time-slot \\
		\hline
		
		$x_t$       & Number of requests arriving in time-slot $t$\\
		\hline
		$r_t$   & Indicator variable; 1 if the service is hosted\\
		&   in time-slot $t$ and 0 otherwise \\
		\hline
		$(x_l - \kappa)^+$ & $\max \{x_l - \kappa, 0\}$\\
		\hline
	\end{tabular}
	\vspace*{10pt}
	\caption{Notation used in Algorithms \ref{algo:RR}}
	\label{table:RR}
\end{table}

\begin{algorithm}
	\caption{RetroRenting (RR)}\label{algo:RR}
	\SetAlgoLined
	
	Input: Fetch cost $M$ units, maximum number of our service requests served by edge server($\kappa$),  rent cost sequence:$\{c_l\}_{l=0}^{t}$, request arrival sequence: $\{x_l\}_{l=0}^t$, $t > 0$\\
	Output:  Service hosting strategy $r_{t+1}$, $t > 0$\\
	Initialize:  Service hosting variable $r_1= t_{\text{fetch}} = t_{\text{evict}} = 0$\\
	\For {\textbf{each} time-slot $t$}{
		$r_{t+1}=r_t$\\
		
		\If{$r_t=0$ }{
			\For{$t_{\text{evict}} < \tau < t$}{
				\If{$\displaystyle\sum_{l=\tau}^t x_l\geq  \displaystyle\sum_{l=\tau}^{t}  c_l+M + \displaystyle\sum_{l=\tau}^t (x_l-\kappa)^+,$}{
					$r_{t+1}=1$, $t_{\text{fetch}} =  t$\\
					break\\
				}			
			}
		}
		\If{$r_t=1$ }{ 
			\For{$t_{\text{fetch}} < \tau < t$}{
				\If{$\displaystyle\sum_{l=\tau}^t x_l + M \leq  \displaystyle\sum_{l=\tau}^{t}  c_l +  \displaystyle\sum_{l=\tau}^t (x_l-\kappa)^+,$}{
					$r_{t+1}=0$, $t_{\text{evict}} = t$\\
					break\\
				}			
			}						
		}
		
	}
\end{algorithm}

\begin{remark}
	Note that in time-slot $t$, the computation and storage complexities of the RR policy scale as $\text{O}(t)$ (if either $t_{\text{fetch}} = 0$ or $t_{\text{evict}} = 0$). This is indeed a limitation of the RR policy since, in the worst case, the computational and storage complexities increase linearly with time. 
	An efficient implementation of RR  is presented in the Appendix. The key idea behind this implementation is, in each time-slot we compute a metric that captures the trend of difference between cost of forwarding and cost of renting. Decisions on when to rent edge resources can be made solely based on this metric.
	This approach is inspired from  \cite{lu2012online}.  Computational and storage complexities of this implementation do not scale with time, thus alleviating the need for the $\text{RR}_u$ policy proposed in \cite{narayana2020retrorenting}. 
\end{remark}

\subsection{Performance guarantees for RR}
\label{subsec:performance_guarantees}

\subsubsection{Adversarial setting}
Our first theorem characterizes the performance of RR in the adversarial  setting.
\begin{theorem}
	\label{thm:RR_adv}
	Let $\rho^{\text{RR}}$ be the competitive ratio of RR. Then,  $\rho^{\text{RR}}\leq\left(4+\frac{2(\kappa-c_{\text{min}})}{M}-\frac{3c_{\text{min}}}{\kappa}\right).$ 
\end{theorem}

Since this result holds for all finite request arrival sequences  and fine rent cost sequences, Theorem \ref{thm:RR_adv} provides a worst-case guarantee on the performance of the RR policy as compared to that of the optimal offline policy. Recall that unlike the RR policy, the optimal offline policy knows the entire arrival sequence and  the entire rent cost sequence  a-priori. 

The competitive ratio of RR improves as the fetch cost ($M$)  increases,  however, it increases linearly with $\kappa.$ Our next result shows that the competitive ratio of \emph{any} deterministic online policy increases linearly  with $\kappa$. 

\begin{theorem}
	\label{thm:any_online}
	Let $\mathcal{P}$ be any deterministic online policy and 
	let $\rho^{\mathcal{P}}$ be the competitive ratio of this policy. 
	Then,
	\begin{align*}
	\rho^{\mathcal{P}} \geq
	\begin{cases*}
	1 + \dfrac{\kappa}{c_{\text{min}} + M} \hspace{1em} \text{ if } \kappa \geq \dfrac{c_{\text{min}}(c_{\text{min}}+M)}{M} \\
	\dfrac{\kappa}{c_{\text{min}}} \hspace{4.8em} \text{ otherwise. }
	\end{cases*}
	\end{align*}
\end{theorem}

From Theorems \ref{thm:RR_adv} and \ref{thm:any_online}, we conclude that the RR policy is order optimal with respect to the edge server computation constraint ($\kappa$) for the setting considered. This is one of the key results of this work. While Theorem \ref{thm:RR_adv} gives a worst-case guarantee on the performance of the RR policy, in our subsequent analytical and simulation results, we observe that for the request sequences considered, the performance of the RR policy is significantly closer to that of the offline optimal policy than the bound in Theorem \ref{thm:RR_adv} suggests. 

\subsubsection{Stochastic arrivals}
Next we characterize the performance of the RR  policy for i.i.d. stochastic arrivals  and negatively associated rent cost sequence (Assumption~\ref{assum_stochastic}, Section \ref{sec:setting}). Recall that, under Assumption \ref{assum_stochastic}, in each time-slot, the number of request arrivals is $x$ with probability $p_x$ for $x = 0, 1, \cdots$.

Our next lemma characterizes the difference between the expected cost incurred in a time-slot by our policies and the optimal online policy.

\begin{lemma}
	\label{lemma:difference_RRstochastic}
	Let $\Delta_t^{\mathcal{P}} = \mathbb{E}[C_t^{\mathcal{P}} - C_t^{\text{OPT-ON}}]$, 
	$\mu = \mathbb{E}[\min\{X_t,\kappa\}]$, $c= \mathbb{E}[R_t]$
 \begin{align*}
	&f(\kappa,\lambda,M,\mu,c) =    (M+\mu) \bigg(\bigg\lceil\frac{\lambda M}{\mu-c}\bigg\rceil \\
	&\frac{\exp\left(-2\frac{(\mu-c)^2 \frac{M}{c}}{(\kappa+c_{\text{max}}-c_{\text{min}})^2}\right)}{1-\exp\left(-2\frac{(\mu-c)^2}{(\kappa+c_{\text{max}}-c_{\text{min}})^2}\right)} \nonumber +\exp\left(-2\frac{(\lambda-1)^2M(\mu-c)}{\lambda (\kappa+c_{\text{max}}-c_{\text{min}})^2}\right)\bigg),\text{ and}\\
	&g(\kappa,\lambda,M,\mu,c) = (c+M) \bigg(2\bigg\lceil\frac{\lambda M}{c-\mu}\bigg\rceil \\
	&\frac{\exp\left(-2\frac{(c-\mu)^2 \frac{M}{\kappa-c}}{(\kappa+c_{\text{max}}-c_{\text{min}})^2}\right)}{1-\exp\left(-2\frac{(c-\mu)^2}{(\kappa+c_{\text{max}}-c_{\text{min}})^2}\right)} + \exp\left(-2\frac{(\lambda-1)^2(c-\mu)M}{\lambda (\kappa+c_{\text{max}}-c_{\text{min}})^2}\right)\bigg).
	\end{align*}
	Then, under Assumption \ref{assum_stochastic},
	\begin{itemize}
		\item[--] Case $\mu > c$:
		\begin{align*}
			\Delta_t^{\text{RR}} &\leq \min_{\lambda: \lambda>1 \text{ and } t > \big\lceil\frac{\lambda M}{\mu-c}\big\rceil}f(\kappa,\lambda,M,\mu,c).
		\end{align*}
		\item[--] Case $\mu < c$:
\begin{align*}
\Delta_t^{\text{RR}} &\leq \min_{\lambda: \lambda>1 \text{ and } t > \big\lceil\frac{\lambda M}{c-\mu}\big\rceil}g(\kappa,\lambda,M,\mu,c).
\end{align*}
	\end{itemize}
\end{lemma}

We thus conclude that for $t$ large enough, the difference between the cost incurred by RR and the optimal online policy in time-slot $t$, decays exponentially with $M$ and $|\mu - c|$.

\begin{theorem}\label{thm:RR_stochastic_theorem}
		Let 
	$\nu = \mathbb{E}[X_t],$ 
	$\mu = \mathbb{E}[\min\{X_t,\kappa\}]$ and  $c= \mathbb{E}[R_t]$.
	Recall the definition of $\sigma^{\mathcal{P}}_T$ given in \eqref{eq:efficiencyRatio}.

		\begin{itemize}
		\item[--] Case $\mu > c$: For the function $f$ defined in Lemma \ref{lemma:difference_RRstochastic},
		\begin{align*}
		\sigma^{\text{RR}}(T)& \leq \min_{\lambda>1} \  \bigg(1 -\dfrac{\big\lceil \frac{\lambda M}{\mu-c}\big\rceil \big(\frac{M+c+\nu}{c + \nu - \mu} + 1\big)}{T} +\\ &\dfrac{T-\big\lceil \frac{\lambda M}{\mu-c}\big\rceil}{T(c+\nu-\mu)}f(\kappa,\lambda,M,\mu,c)\bigg),
			\end{align*}

		\item[--] Case $\mu < c$: For the function $g$ defined in Lemma \ref{lemma:difference_RRstochastic},
		\begin{align*}
		\sigma^{\text{RR}}(T) &\leq \min_{\lambda>1} \  \bigg(1 -\dfrac{\big\lceil \frac{\lambda M}{c-\mu}\big\rceil \big(\frac{M+c+\nu}{\nu} + 1\big)}{T} +\\
		& \dfrac{T-\big\lceil \frac{\lambda M}{c-\mu}\big\rceil}{T \nu}g(\kappa,\lambda,M,\mu,c)\bigg),
		\end{align*}
		
	\end{itemize}
\end{theorem}

\begin{remark}
	The bounds obtained in Lemma \ref{lemma:difference_RRstochastic} and Theorem \ref{thm:RR_stochastic_theorem} hold for all i.i.d. stochastic arrival processes  and negatively associated rent costs. We note that the bounds worsen as $\kappa$ increases. This is a consequence of using Hoeffding's inequality to bound the probability of certain events. It is important to note that significantly tighter bounds can be obtained for specific i.i.d. processes by using the Chernoff bound instead of Hoeffding's inequality. In the next section, via simulations, we show that the performance of RR  does not worsen as $\kappa$ increases. We thus conclude that the deterioration of the performance guarantees with increase in $\kappa$ is a consequence of the analytical tools used and not fundamental to RR.
 Note that for $\mu=c$ the performance of RR is same as that of optimal online policy.
\end{remark}

We use Theorem \ref{thm:RR_stochastic_theorem} to conclude that for $T$ large enough, the bound on the ratio of the expected cost incurred by RR in $T$ time-slots to that of the optimal online policy (OPT-ON) in the same time interval decays as $M$ increases.  

Often, policies designed with the objective of minimizing the competitive ratio tend to perform poorly on average in typical stochastic settings.  Similarly, polices designed for specific stochastic arrival processes can have poor competitive ratios if they perform poorly for certain `corner case' arrival sequences. The performance guarantees for RR obtained in this section show that RR performs well in both the adversarial and the i.i.d. stochastic settings. This is a noteworthy feature of the RR policy.

\subsection{Performance of TTL}\label{subsec:otherpolicies}
In this section, we focus on the TTL policy which is widely used and studied in the classical caching literature. TTL serves as a benchmark to compare the performance of RR. 

The TTL policy fetches and hosts  the service whenever there is a miss, i.e., the service is requested but is not hosted  on the edge server. There is a timer associated with the fetch, which is set to a fixed value ($L$) right after the service is fetched. If the service is not requested before the timer expires, the service is evicted from the host. If a request arrives while the service the hosted, the timer is reset to its initial value of $L$. Refer to Algorithm \ref{algo:TTLfixed} for a formal definition.

\begin{algorithm}
  \caption{TTL Policy}\label{algo:TTLfixed}
 \SetAlgoLined
Input: Number of request arrived $x_t, \ t>0$, TTL value $L$\\
Output: Service hosting strategy $r_{t+1}, \ t>0$ \\
Initialize:  Service hosting variable $r_1=0$ and $\text{timer}=0$ \\
\For{\textbf{each} time-slot $t$}{
\uIf {$x_{t}=0$}{
	\uIf {$\text{timer}=0$}{
		$r_{t+1} = 0$\\
	}
	\Else{
		$r_{t+1} = 1$, $\text{timer} = \text{timer} - 1$
	}
}
\Else{
	$r_{t+1} = 1$, $\text{timer} = L$
}
}

\end{algorithm}

Our next result provides a lower bound on the competitive ratio of the TTL policy. 

 \begin{theorem}\label{thm:fixed_TTL}
Let ${\rho}^{\text{TTL}}$ be the competitive ratio of the TTL policy with TTL value $L$ as defined in \eqref{eq:competitiveRatio}. Then,

\begin{align*}
\rho^{\text{TTL}} \geq
\begin{cases*}
1+Lc_{\text{max}}+M \hspace{3.7em} \text{ if } 1\leq \kappa < M+c_{\text{max}}, \\
\dfrac{\kappa+Lc_{\text{max}}+M}{c_{\text{max}}+\min\{Lc_{\text{max}},M\}} \hspace{0.8em} \text{ otherwise. }
\end{cases*}
\end{align*}
%
\end{theorem} 

The key takeaway from Theorem \ref{thm:fixed_TTL} is that unlike the RR policy, the performance of the TTL policy deteriorates as the fetch cost ($M$) increases. This is a consequence of the fact that the TTL policy fetches and hosts  the service on a miss irrespective of the value of $M$, whereas for high values of $M$, RR and the optimal offline policy might choose not to fetch the service at all.
Note that the performance of both RR and TTL deteriorates with increase in $\kappa$. 

Next, we characterize the performance of TTL for i.i.d. stochastic arrivals  and negatively associated rent costs. 

\begin{lemma}
	\label{lemma:TTLstochastic}
		Let $\Delta^{TTL}_t = \mathbb{E}[C_t^{TTL} - C_t^{\text{OPT-ON}}]$, $\mu = \mathbb{E}[\min\{X_t,\kappa\}]$ and $c= \mathbb{E}[R_t]$. Recall that $p_0 = \mathbb{P}(X_t = 0)$. Then, under Assumption \ref{assum_stochastic}, if $\mu<c$,
		\begin{align*}
		\Delta^{TTL}_t =& p_0^{\min\{t-1,L\}} (1-p_0)M  +  \left(1-p_0^{\min\{t-1,L\}}\right) (c - \mu) .
		\end{align*}
		For $t>1$, $\min\{t-1,L\} \geq 1$, and therefore, 
		\begin{align*}
		\Delta^{TTL}_t \geq& \min\{(1-p_0)(p_0M + c - \mu), c - \mu\}.
		\end{align*}
\end{lemma}

We thus conclude that for low arrivals rates, the difference between the cost incurred by TTL and the optimal online policy in time-slot $t$, increases with $M$ and $c-\mu$. This illustrates the limitations of TTL for the stochastic setting with low arrival rates. The sub-optimality of TTL is a consequence of the fact that the TTL policy fetches and hosts  the service on a miss irrespective of the value of $M$ and $\mu$, whereas for high values of $M$ and low values of $\mu$, RR and the optimal online policy choose not to fetch the service at all.

\begin{theorem}\label{thm:TTL_stochastic_theorem}
	Let 
	$\nu = \mathbb{E}[X_t]$, 
	$\mu = \mathbb{E}[\min\{X_t,\kappa\}]$ and $c= \mathbb{E}[R_t]$. Recall the definition of $\sigma^{\mathcal{P}}_T$ given in \eqref{eq:efficiencyRatio}.
	If $\mu < c$,
		\begin{align*}
		\sigma^{\text{TTL}}(T) \geq \left(1-\dfrac{1}{T} \right)\dfrac{\min\{(1-p_0)(p_0M + c - \mu), c - \mu\}}{\nu}.
		\end{align*}
\end{theorem}


From Theorem \ref{thm:TTL_stochastic_theorem}, we conclude that for low request arrival rates ($\mu < c$), the performance of TTL deteriorates with increases in $M$. Contrary to this, the performance of RR  approaches the performance of the optimal online policy as $M$ increases (Theorem \ref{thm:RR_stochastic_theorem}). We thus conclude that for low arrival rates and high fetch cost, TTL is sub-optimal. 

TTL policies perform well in content caching, where on a miss, the requested content is fetched from the back-end server by all policies including TTL. However, as discussed above, on a miss in our setting, there are two options: \textit{(a) request forwarding} which forwards the  request to the back-end server for service; and \textit{(b) service fetch} which fetches all the data and code needed for running the service from the back-end server and hosts  it on the edge server. The cost for these two actions is different. By definition, TTL always takes the second option, whereas, for low request arrival rates and high fetch cost, RR, and the optimal online and offline policies use the first option. This explains the poor performance of TTL for our setting.

\section{Simulation Results}
\label{sec:simulation}
In this section, we compare the performance of various policies via simulations.  
From our analytical results, we know that the Fixed TTL policy performs poorly for our setting. Therefore, we compare our policy with an online variant of the TTL policy proposed in \cite{carra2019ttl}.
The simulation parameters for each set of simulation are provided in the figure caption.

\subsection{Stochastic Arrivals and Stochastic Rent Costs}

For the first set of simulations, we consider stochastic arrival and rent processes, and use an i.i.d. Bernoulli sequence for determining the arrivals in each slot. We model the time-varying rent cost sequence using the Auto-Regressive Moving-Average (ARMA) model \cite{box2011time} and to choose the model's hyperparameters, we fit it to real-world spot prices of spare server capacity in the  AWS cloud as given in  \cite{awsprices}. We collect the chronological spot prices data for the instance 'm4.large' in the Central Canada region and employed the grid search method \cite{Arma_grid} on this data to estimate the hyperparameters of the ARMA model that minimize the mean square error. Finally, we add a constant shift to make sure that all cost values are positive and do a normalization to get a desired average cost value.

We compare the performance of RR, TTL online, and the optimal offline policy. In addition to these, we plot the lower bound on the cost incurred by any online policy (Lemma \ref{lemma:optimal_causal}). Each data-point in the plots is averaged over 10000 time-slots. 

\subsubsection{i.i.d. Bernoulli Arrivals}
\begin{figure}
    \centering
    \begin{minipage}{0.45\textwidth}
        \begin{center}
		\includegraphics[width=\linewidth]{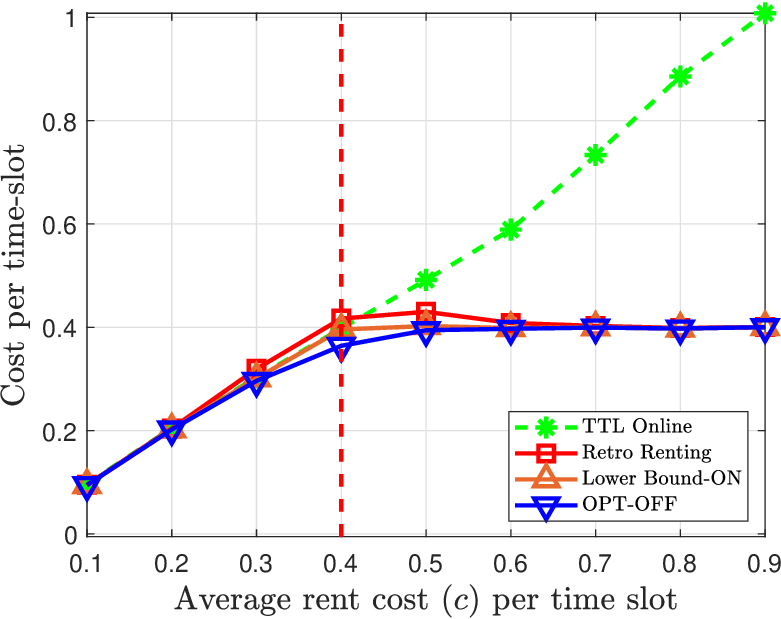}
		\caption{Cost per time-slot as a function of average rent cost ($c$) for $M = 4$ for i.i.d. Bernoulli$(p)$ arrivals with $p = 0.4$}
		\label{fig:RR_c}
	     \end{center}
    \end{minipage}\hfill
    \begin{minipage}{0.45\textwidth}
        \begin{center}
		\includegraphics[width=\linewidth]{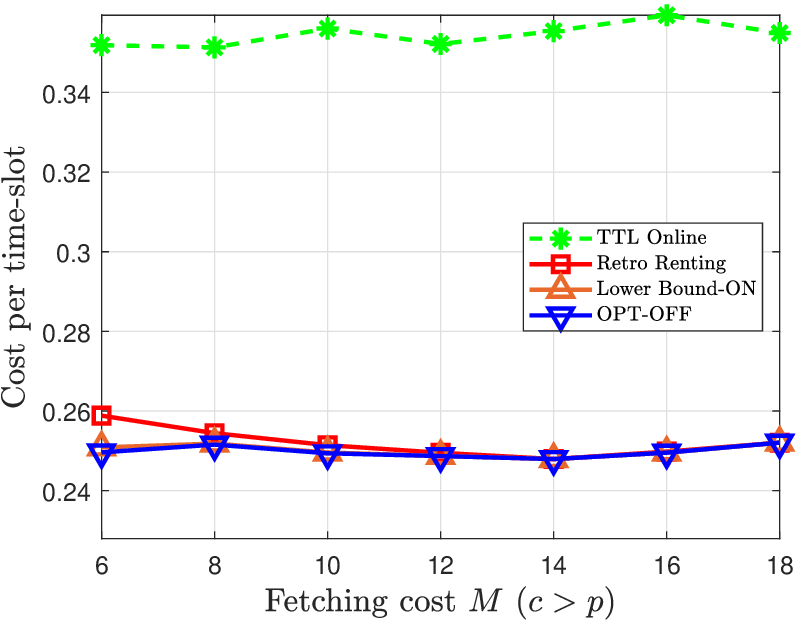}
		\caption{Cost per time-slot as a function of the fetch cost ($M$) for i.i.d. Bernoulli$(p)$ arrivals with $p = 0.25$ and  ARMA(4,2) rent costs  with mean $c=0.35$}
	\end{center}
	\label{fig:RR_M_cgtp}
    \end{minipage}
\end{figure}

In Figures \ref{fig:RR_c}-\ref{fig:RR_M_pgtc}, we consider Bernoulli request arrivals with parameter $p$, i.e., $X_t = 1$ with probability $p$ and $X_t = 0$ otherwise. Recall that $\kappa \geq 1$ and $\mu = \mathbb{E}[\min\{X_t, \kappa\}]$. Since $X_t \leq 1$ in this case, therefore, $\mu = p$. We compare the performance of RR with the optimal offline and TTL online policies. 

 The performance of the RR policy is quite close to that of the optimal online policy for all parameter values considered. The performance gap between the optimal offline policy and the RR policy is  small compared to the bound on competitive ratio obtained in Theorem \ref{thm:RR_adv}. We see that the gap between the performance of the RR and optimal online policy increases as $M$ and/or $|\mu-c|$ decrease. This can be explained as follows. If $\mu<c$, the optimal online policy does not fetch/store the service and forwards all the requests to the back-end server. However, for small values of $c-\mu$, and $M$, the condition the RR policy checks to fetch and host  the service (Step 8 in Algorithm \ref{algo:RR}) is not very unlikely. This leads to multiple fetch--store--evict cycles and therefore a higher cost than the optimal online policy. As $M$ and/or $c-\mu$ increase, this event becomes less probable. The case when $\mu>c$ can be argued along similar lines. 	

\subsubsection{i.i.d. Poisson Arrivals}
\begin{figure}
    \centering
    \begin{minipage}{0.45\textwidth}
        	\begin{center}
		\includegraphics[width=\linewidth]{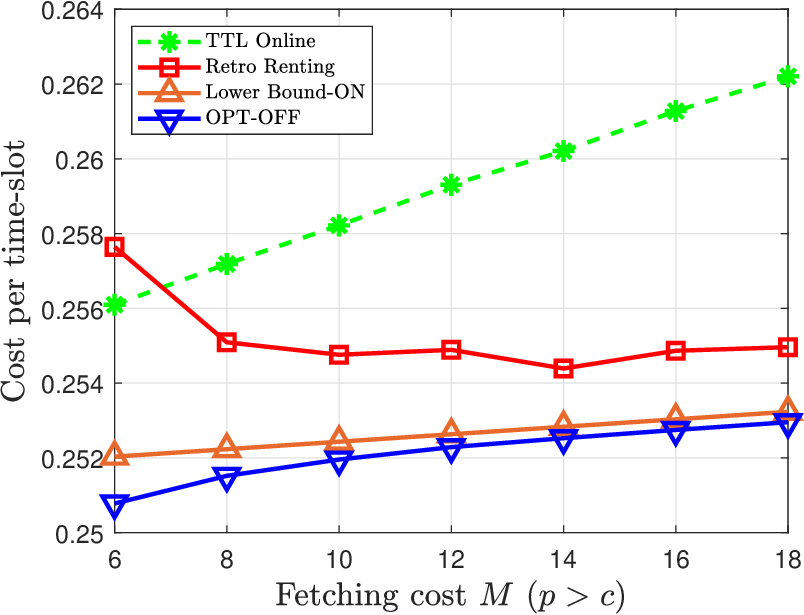}
		\caption{Cost per time-slot as a function of the fetch cost ($M$) for i.i.d. Bernoulli$(p)$ arrivals with  $p = 0.35$  and ARMA(4,2) rent costs  with mean $c=0.25$}
		\label{fig:RR_M_pgtc}
	\end{center}
	
    \end{minipage}\hfill
    \begin{minipage}{0.45\textwidth}
        	\begin{center}
		\includegraphics[width=\linewidth]{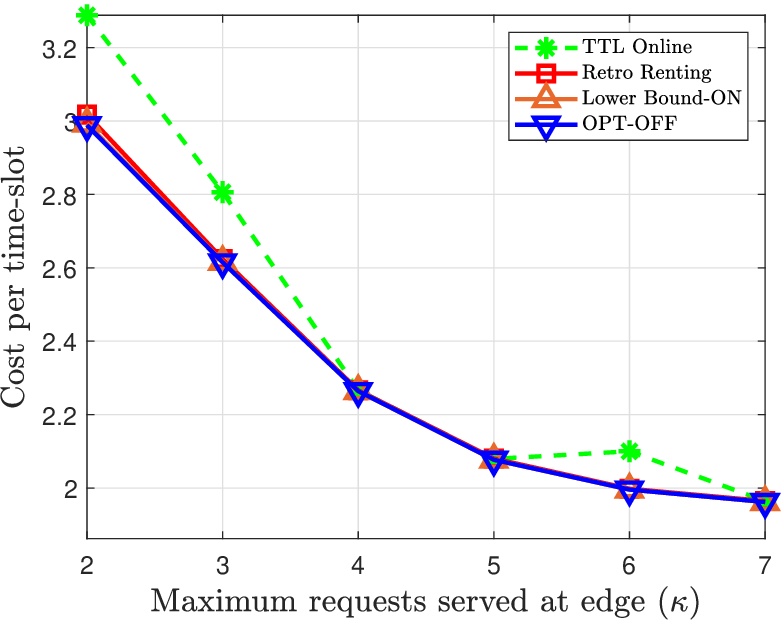}
		\caption{Cost per time-slot as a function of  ($\kappa$) for i.i.d. Poisson arrivals with parameter $\lambda=3$, $M=20$, and  ARMA(4,2) rent costs  with mean $c=1.8$}
		\label{fig:RR_Poisson_kappa}
	\end{center}
    \end{minipage}
\end{figure}

In Figure \ref{fig:RR_Poisson_kappa}, we consider the case where the arrival process is Poisson with parameter $\lambda$. We vary $\kappa$. We see that the performance of all policies improves with increase in $\kappa$. The performance of RR is very close that of the optimal offline policy and the lower bound on online policies. 

\begin{figure}
    \centering
    
    \begin{minipage}{0.45\textwidth}
        \begin{center}
		\includegraphics[width=\linewidth]{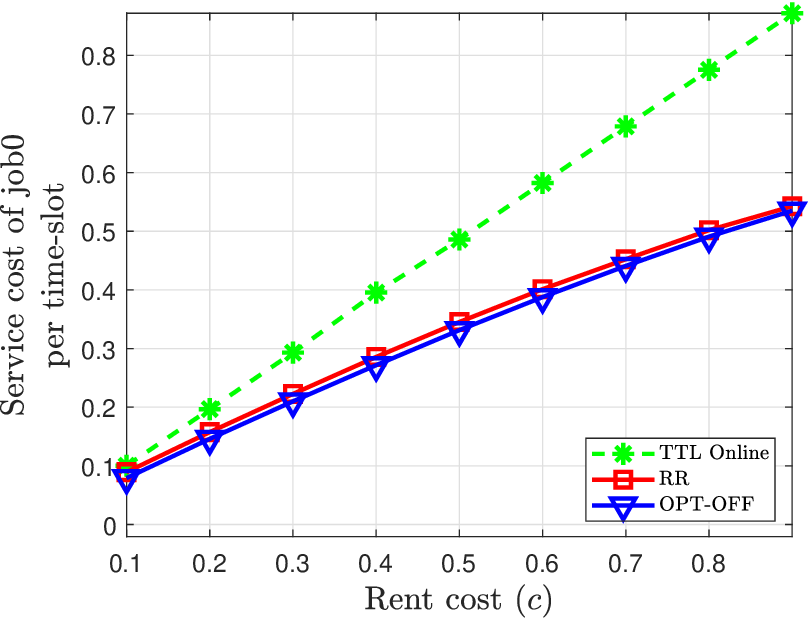}
		\caption{Cost per time-slot as a function of storage cost ($c$) for $M = 10$ for Job 0}\label{fig:job0_c}
	\end{center}
    \end{minipage}\hfill
    \begin{minipage}{0.45\textwidth}
        	\begin{center}
		\includegraphics[width=\linewidth]{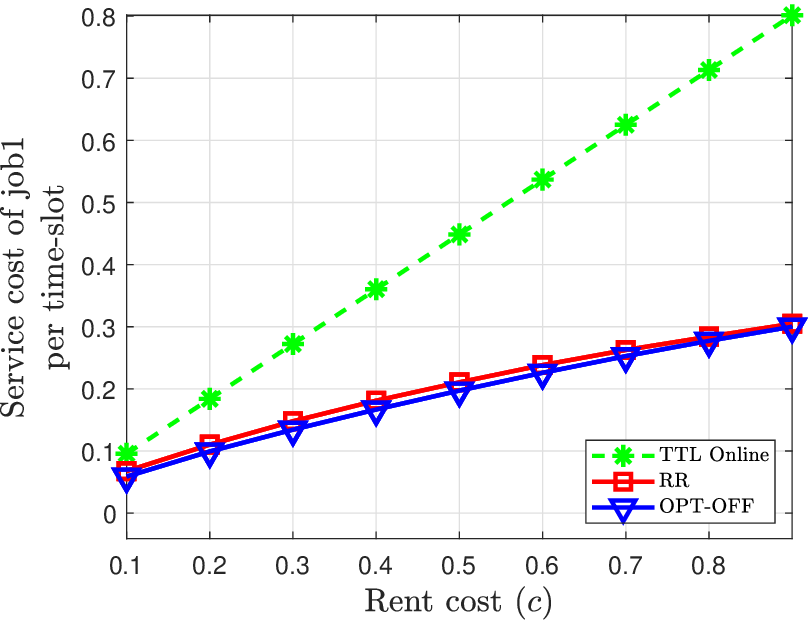}
		\caption{Cost per time-slot as a function of storage cost ($c$) for $M = 10$ for Job 1}\label{fig:job1_c}
	\end{center}
    \end{minipage}
\end{figure}

\begin{figure}
    \centering
    \begin{minipage}{0.45\textwidth}
        	\begin{center}
		\includegraphics[width=\linewidth]{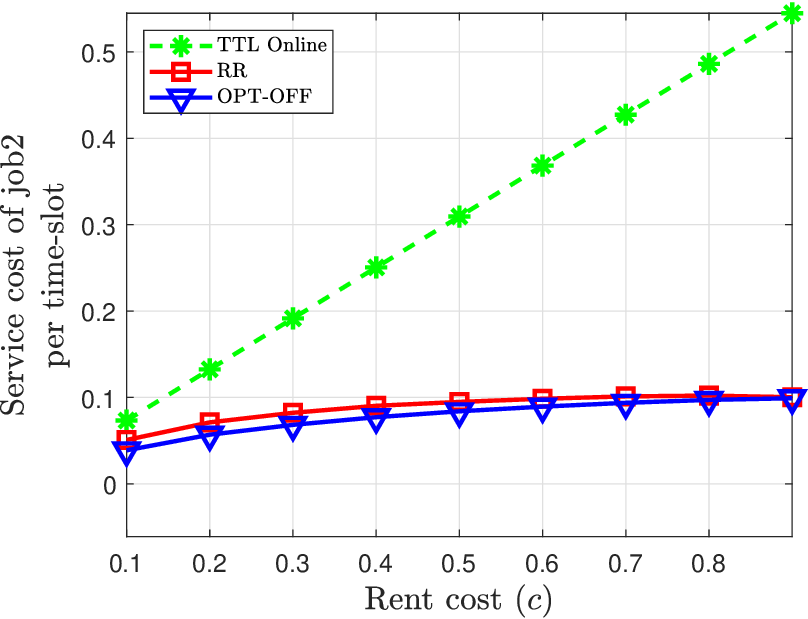}
		\caption{Cost per time-slot as a function of storage cost ($c$) for $M = 10$ for Job 2}\label{fig:job2_c}
	\end{center}
    \end{minipage}\hfill
    \begin{minipage}{0.45\textwidth}
        \begin{center}
		\includegraphics[width=\linewidth]{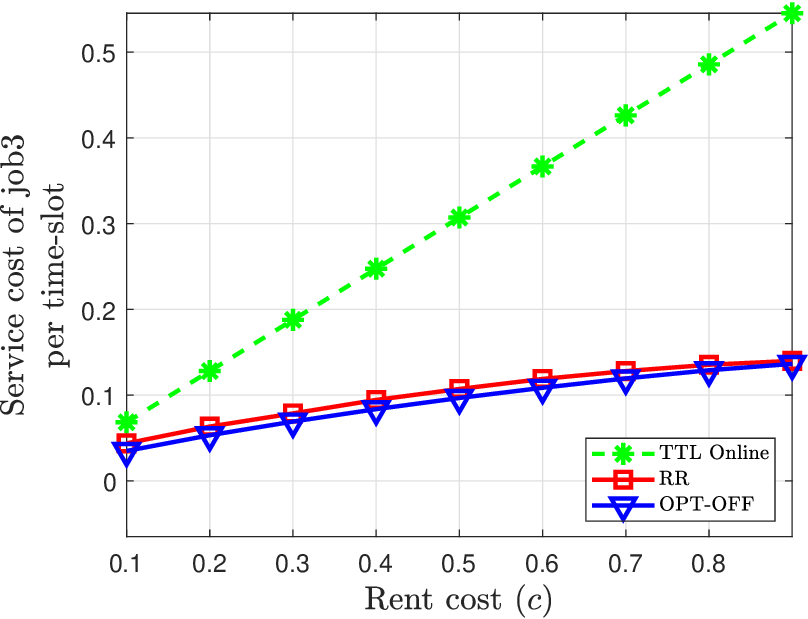}
		\caption{Cost per time-slot as a function of storage cost ($c$) for $M = 10$ for Job 3}\label{fig:job3_c}
	\end{center}
    \end{minipage}
\end{figure}

\begin{figure}
    \centering
    \begin{minipage}{0.45\textwidth}
        	\begin{center}
		\includegraphics[width=\linewidth]{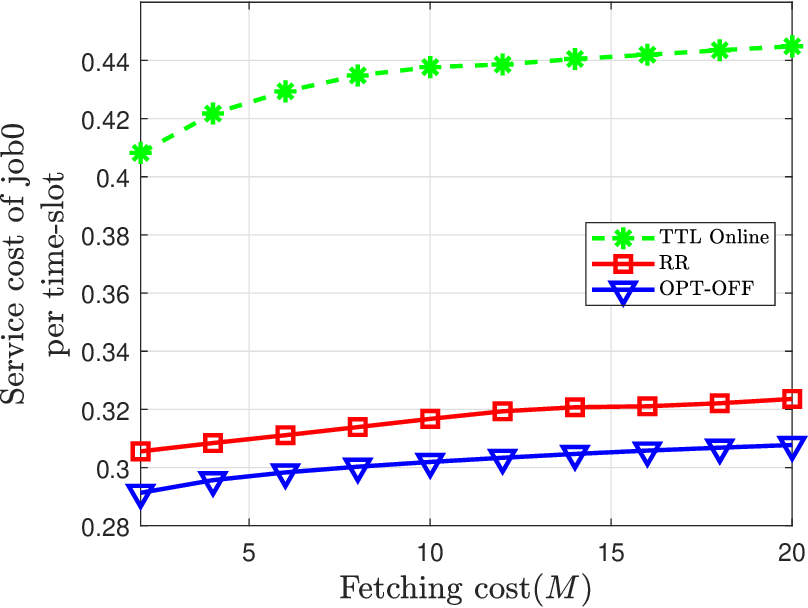}
		\caption{Cost per time-slot as a function of the fetch cost ($M$) for $c=0.45$ for Job 0}\label{fig:job0_M}
	\end{center}
    \end{minipage}\hfill
    \begin{minipage}{0.45\textwidth}
        \begin{center}
		\includegraphics[width=\linewidth]{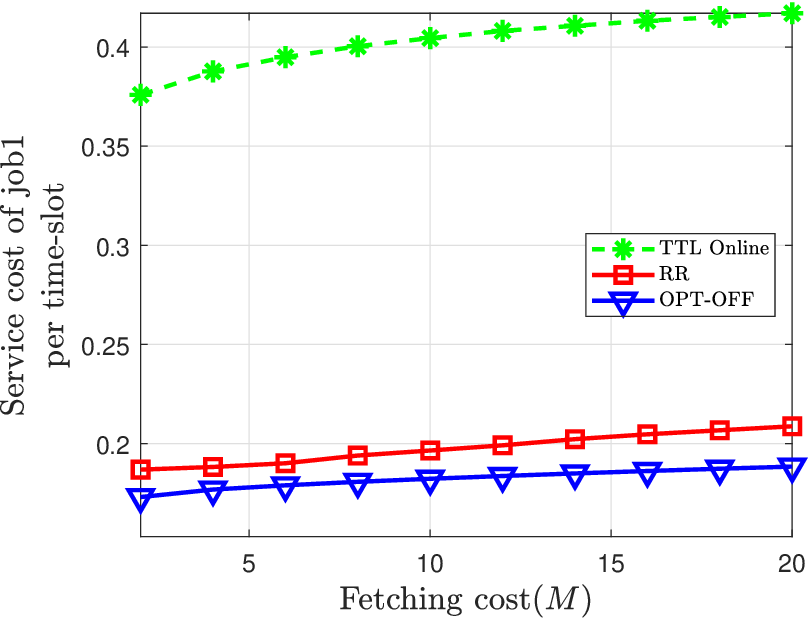}
		\caption{Cost per time-slot as a function of the fetch cost ($M$) for $c=0.45$ for Job 1}\label{fig:job1_M}
	\end{center}
    \end{minipage}
\end{figure}

\subsection{Trace-driven Simulations}

\begin{figure}
    \centering
    \begin{minipage}{0.45\textwidth}
        	\begin{center}
		\includegraphics[width=\linewidth]{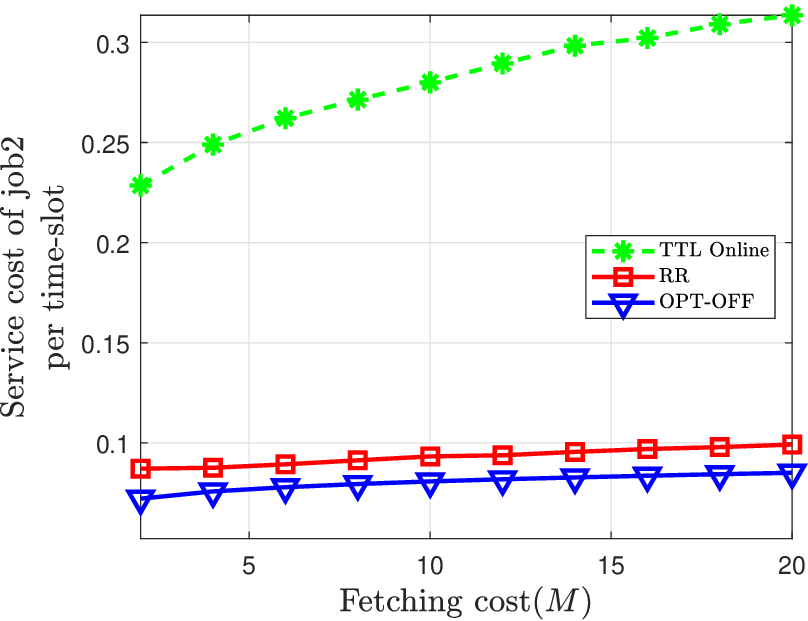}
		\caption{Cost per time-slot as a function of the fetch cost ($M$) for $c=0.45$ for Job 2}\label{fig:job2_M}
	\end{center}
    \end{minipage}\hfill
    \begin{minipage}{0.45\textwidth}
        	\begin{center}
		\includegraphics[width=\linewidth]{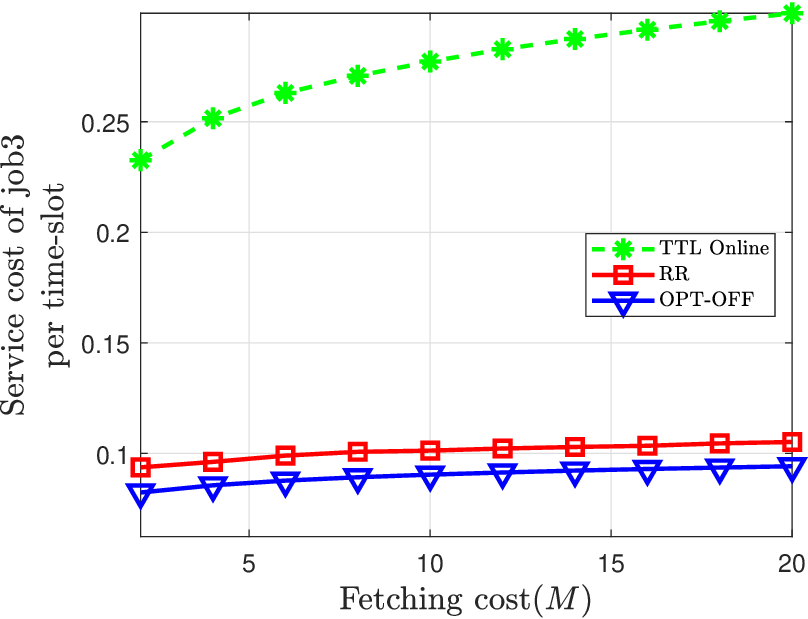}
		\caption{Cost per time-slot as a function of the fetch cost ($M$) for $c=0.45$ for Job 3}\label{fig:job3_M}
	\end{center}
    \end{minipage}
\end{figure}
\begin{figure}
    \centering
    \begin{minipage}{0.45\textwidth}
        \begin{center}
		\includegraphics[width=\linewidth]{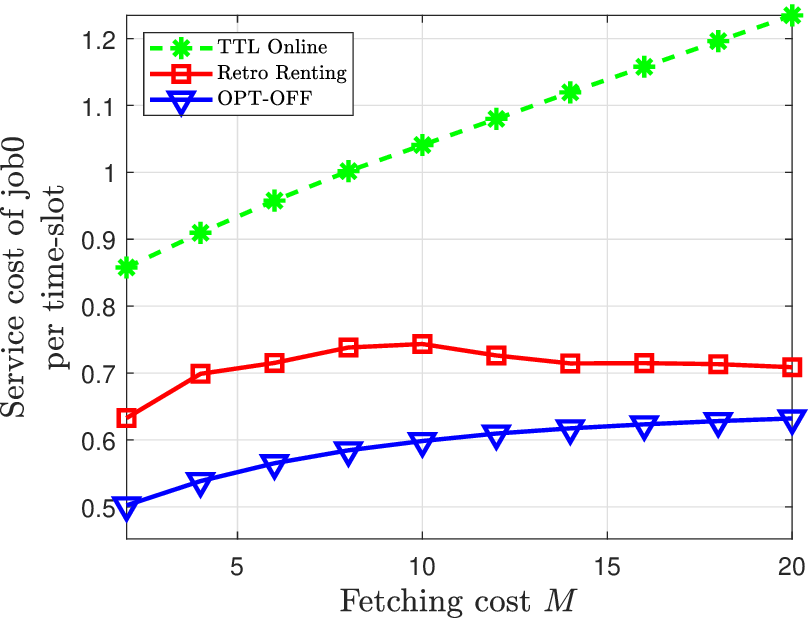}
		\caption{Cost per time-slot as a function of the fetch cost ($M$) for time varying renting cost for Job 0}\label{fig:advcost_job0_M}
	\end{center}
    \end{minipage}\hfill
    \begin{minipage}{0.45\textwidth}
        \begin{center}
		\includegraphics[width=\linewidth]{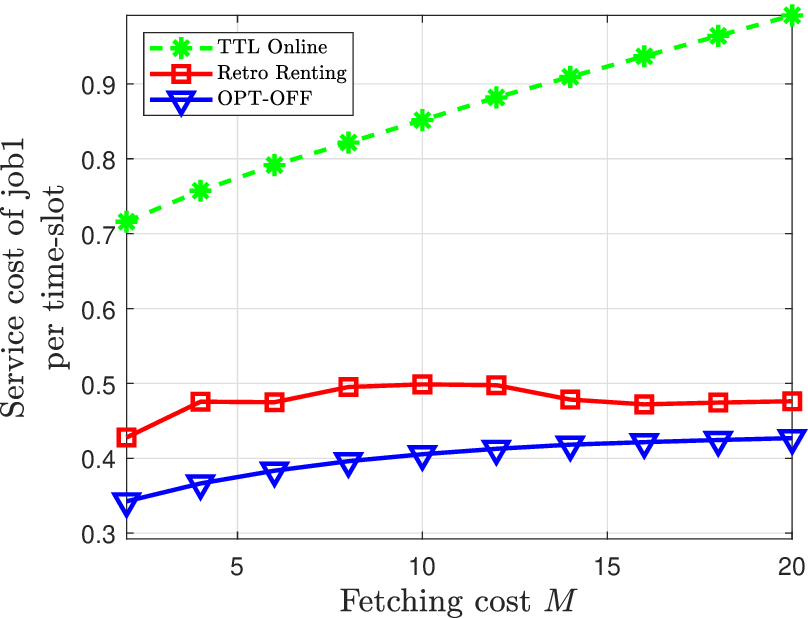}
		\caption{Cost per time-slot as a function of the fetch cost ($M$) for time varying renting cost for Job 1}\label{fig:advcost_job1_M}
	\end{center}
    \end{minipage}
\end{figure}

For the next set of simulations, we use trace-data obtained from a Google Cluster \cite{googlecluster} for the arrival process. We use a time-slot duration small enough to ensure that there is at most one request in a time-slot. This trace-data has requests for four types of jobs/services identified as ``Job 0", ``Job 1", ``Job 2", and ``Job 3". In this section, we present  results for ``Job 0" (Figures \ref{fig:job0_c}, \ref{fig:job0_M}, \ref{fig:advcost_job0_M} ), ``Job 1" (Figures \ref{fig:job1_c}, \ref{fig:job1_M}, \ref{fig:advcost_job1_M}), ``Job 2" (Figures \ref{fig:job2_c} and \ref{fig:job2_M}), and ``Job 3" (Figures \ref{fig:job3_c}, \ref{fig:job3_M}). 
In Figures \ref{fig:job0_c}-\ref{fig:job3_M}, we use the rent cost $c_t$ to be constant across time-slots. Whereas in Figures \ref{fig:advcost_job0_M}, \ref{fig:advcost_job1_M} we use time-varying rent cost obtained from real world data. Specifically, we use the time-varying spot prices of spare server capacity in the  AWS cloud as given in \cite{awsprices}. This data \cite{awsprices} consists of details like current spot price for different instances for different regions. We use the spot price details of Canada region for one instance 'm4.large'. We normalise this data by dividing each entry by average value of spot prices. So the empirical of new cost cost data is equal to one.

We compare the performance of RR  with the optimal offline policy and TTL online.
The performance of TTL online is the worst among these polices, and the performance gap between TTL online and RR is significant.

\section{Proof Outlines}
\label{sec:proofOutlines}
\subsection{Proof Outline for Theorem \ref{thm:RR_adv}}
We divide time into frames such that Frame $i$ for $i \in \mathbb{Z}^+$ starts when OPT-OFF downloads the service for the $i^{\text{th}}$ time. We refer to the time interval before the beginning of the first frame as Frame 0. Note that by definition, in all frames, except maybe the last frame, there is exactly one eviction by OPT-OFF.

\begin{figure}[ht]
	\centering
	\begin{tikzpicture}
	\foreach \x in {0,0.3,0.6,...,7.2}{
		\draw[] (\x,0) --  (\x+0.3,0);
		\draw[gray] (\x,-1mm) -- (\x,1mm);
	}
	
	\draw[<->,color=black] (0.25,14mm) -- 
	node[above=0mm,pos=0.5]{$i.a$} (2.35,14mm);
	\draw[<->,color=black] (2.35,14mm) -- 
	node[above=0mm,pos=0.5]{$i.b$} (4.15,14mm);
	\draw[<->,color=black] (4.15,14mm) -- 
	node[above=0mm,pos=0.5]{$i.c$} (5.65,14mm);
	\draw[<->,color=black] (5.65,14mm) -- 
	node[above=0mm,pos=0.5]{$i.d$} (6.85,14mm);
	
	\foreach \x in {0.25,2.35,4.15,5.65,6.85}{
		\draw[gray] (\x,12mm) -- (\x,16mm);
	}
	\draw[gray] (7.2,-1mm) -- (7.2,1mm);
	\draw[-latex] (0.25,10mm) -- node[above=5mm]{} (0.25,0mm);
	\draw[-latex] (6.85,10mm) -- node[above=5mm]{} (6.85,0mm);
	\draw[-latex] (4.15,0mm) -- node[above=5mm]{} (4.15,10mm);
	\draw[-latex,color=red] (2.35,10mm) -- node[above=5mm]{} (2.35,0mm);
	\draw[-latex,color=red] (5.65,0mm) -- node[above=5mm]{} (5.65,10mm);
	
	\draw[<->,color=black] (0.3,-2mm) -- node[below=0mm,pos=0.5]{Frame 
		$i$} (6.9,-2mm);
	
	\draw[black] (0.3,-1mm) -- (0.3,-3mm);
	\draw[black] (6.9,-1mm) -- (6.9,-3mm);
	
	\node[] at (7.65,-0.85) {OPT-OFF};
	
	\filldraw[fill=black!] (0.25,-1) rectangle (4.15,-0.7);
	\filldraw[fill=white!40] (4.15,-1) rectangle (6.85,-0.7);
	\node[] at (7.25,-1.4) {RR};
	\filldraw[fill=white!40] (0.25,-1.5) rectangle (2.4,-1.2);
	\filldraw[fill=red!] (2.4,-1.5) rectangle (5.7,-1.2);
	\filldraw[fill=white!40] (5.7,-1.5) rectangle (6.85,-1.2);

	\end{tikzpicture}
	\caption{Illustration showing 
	download and eviction by OPT-OFF and  RR in the $i^{\text{th}}$ frame. 
	Downward arrows represent fetches, upward arrows indicate evictions. 
	Black and red arrows correspond to the OPT-OFF and RR policy respectively. 
	The two bars below the time-line indicate the state of the system under 
	OPT-OFF and  RR. The solid black and solid red portions represent the 
	intervals during with OPT-OFF and RR host  the service respectively.}
\label{fig:OPT-OFF_RR_frame_proofoutline}
\end{figure}
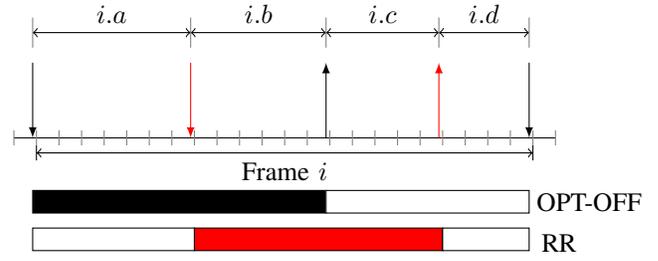

We use the properties of RR and OPT-OFF to show that each frame in which OPT-OFF evicts the service has the following structure (Figure \ref{fig:OPT-OFF_RR_frame_proofoutline}):
\begin{enumerate}
	\item[--] RR fetches and evicts the service exactly once each. 
	\item[--] RR does not host  the service at the beginning of the frame.
    \item[--] The fetch by RR in Frame $i$ is before OPT-OFF evicts the service in Frame $i$.
    \item[--] The eviction by RR in Frame $i$ is after OPT-OFF evicts the service in Frame $i$.
\end{enumerate}
We note that both RR and OPT-OFF fetch exactly once in a frame and therefore, the fetch cost under RR and OPT-OFF is identical for both policies. 

We divide Frame $i$ into four sub-frames defined as follows. 
\begin{enumerate}
	\item[--] $i.a$: OPT-OFF hosts  the service, while RR does not.
    \item[--] $i.b$: OPT-OFF and RR both host  the service.
    \item[--] $i.c$: OPT-OFF does not host  the service while RR does.
    \item[--] $i.d$: OPT-OFF and RR both don't host  the service. 
\end{enumerate}

The service and rent costs are identical for OPT-OFF and RR in Sub-frame $i.b$ and Sub-frame $i.d$. 

We show that the difference between the service and rent costs incurred by RR and OPT-OFF in Sub-frame $i.a$ and Sub-frame $i.c$ is upper bounded by $M+\kappa-c_{\text{min}}$ and $M+c_{\text{max}}$ respectively.

We use the two previous steps to show that the cumulative difference in service and rent costs under RR and OPT-OFF in a frame is upper bounded by $2M+\kappa+c_{\text{max}}-c_{\text{min}}$. 
Since the fetch cost under RR and OPT-OFF in a frame is equal, we have that the total cost incurred by RR and OPT-OFF in a frame differs by at most $2M+\kappa+c_{\text{max}}-c_{\text{min}}$. 

We show that once fetched, OPT-OFF hosts  the service for at least $\frac{M}{\kappa-c_{\text{min}}}$ time-slots (Lemma \ref{lem:OPT_slots}). We thus conclude that the total cost incurred by OPT-OFF in a frame is lower bounded by $M + \frac{Mc_{\text{min}}}{\kappa-c_{\text{min}}} $. We use this to upper bound the ratio of the cost incurred by RR and cost incurred by OPT-OFF in the frame.

The cost incurred by RR and OPT-OFF in Frame 0 is equal. We then focus on the last frame. If OPT-OFF does evict the service in the last frame, the analysis is identical to that of the previous frame. Else, we upper bound the ratio of the cost incurred by RR and cost incurred by OPT-OFF in the frame.

The final result then follows from stitching together the results obtained for individual frames.

\subsection{Proof Outline for Theorem \ref{thm:any_online}}

We divide the class of deterministic online policies into two sub-classes. Any policy in the first sub-class hosts  the service during the first time-slot. All other polices are in the second sub-class. 

For each policy in either sub-class, we construct a specific arrival sequence,   a specific rent cost sequence  and compute the ratio of the cost incurred by the deterministic online policy and an alternative policy. By definition, this ratio serves as a lower bound on the competitive ratio of the deterministic online policy.

\subsection{Proof Outline for Theorem \ref{thm:RR_stochastic_theorem}}

We first characterize a lower bound on the cost per time-slot incurred by any online policy (Lemma \ref{lemma:optimal_causal}).

Next, we focus on the case where $\mu > c$. We upper bound the probability of the service not being hosted during time-slot $t$ under RR  using Hoeffding's inequality \cite{hoeffding1994probability}. The result then follows by the fact that conditioned on the service being hosted in time-slot $t$, the expected total cost incurred by RR  is at most $\mathbb{E}[R_t]+\mathbb{E}[X_t - \min\{X_t, \kappa\}] = c + \mathbb{E}[X_t] - \mu$ and is upper bounded by $M + c + \mathbb{E}[X_t]$ otherwise. 
We then consider the case where $\mu < c$. We upper bound the probability of the service being hosted or being fetched in time-slot $t$ under RR  using Hoeffding's inequality  \cite{hoeffding1994probability}. The result then follows by the fact that conditioned on the service not being hosted and not being fetched in time-slot $t$, the expected total cost incurred by RR is at most $\mathbb{E}[X_t]$ and is upper bounded by $M + c + \mathbb{E}[X_t]$ otherwise. 
\subsection{Proof Outline for Theorem \ref{thm:fixed_TTL}}
Depending on the value of the system parameters ($M$, $c_{\text{min}}$, $c_{\text{max}}$, $\kappa$), we construct specific arrival sequences,  rent cost sequences and compute the ratio of the cost incurred by TTL and OPT-OFF for these  sequences. By definition, this ratio serves as a lower bound on the competitive ratio of TTL in each case.
\subsection{Proof Outline for Theorem \ref{thm:TTL_stochastic_theorem}}
Under the TTL policy with TTL value $L$, the service is hosted during a given time-slot if and only if it is requested at least once in the $L$ previous time-slots. We compute the probability of the event $G$ defined as the event that the service is requested at least once in the $L$ previous time-slots. We compute the conditional expected cost incurred by the TTL policy in time-slot conditioned on $G$ and $G^c$ to obtain the result.
\section{Proofs}
\label{sec:proofs}
\subsection{Proof of Theorem \ref{thm:RR_adv}}
The notation used in this subsection is given in Table \ref{table:proofs}. 
\begin{table}[ht]
	\centering
	\begin{tabular}{ |c | l|} 
		\hline
		\textbf{Symbol}&  \textbf{Description} \\ 
		\hline
		\hline
		$t$ & Time index\\
		\hline
		$M$ & Fetch cost\\
		\hline
		$c_t$       & Rent cost per time-slot  $t$\\
		\hline
		$c_{\text{min}}$       &Minimum value of $c_t$\\
		\hline
		$c_{\text{max}}$       &Maximum value of $c_t$\\
		\hline
		
		$x_t$       & Number of requests arriving in time-slot $t$\\
		\hline
		$\underbar{\text{$x$}}_t$       & $\min \{x_t, \kappa\}$ \\
		\hline
		$\delta_t$       & $x_t-\underbar{\text{$x$}}_t$\\
		\hline
		$r^*(t)$   & Indicator variable; 1 if the service is hosted \\
		& by OPT-OFF during time-slot $t$ and 0 otherwise \\
		\hline
		$\text{  }$  &  $\text{ }$\\
		$r^{\text{RR}}(t)$   & Indicator variable; 1 if the service is hosted\\
		&  by RR during time-slot $t$ and 0 otherwise \\
		\hline
		$\eta$   & a policy\\
		\hline
		$\text{  }$  &  $\text{ }$\\
		$C^{\eta}(n,m)$   & Total cost incurred by the policy $\eta$ in the	 interval $[n,m]$\\
		\hline
		$\text{  }$  &  $\text{ }$\\
		$C^{\text{OPT-OFF}}(n,m)$  & Total cost incurred by the offline optimal \\
		&policy in the interval $[n,m]$\\
		\hline
		Frame $i$  & The interval between the $i^{\text{th}}$ and the \\
		& $(i+1)^{\text{th}}$ fetch by the offline optimal policy\\
		\hline
		$\text{  }$  &  $\text{ }$\\
		$C^{\text{OPT-OFF}}(i)$   & Total cost incurred by the \\
		&offline optimal policy in Frame $i$\\
		\hline
		$\text{  }$  &  $\text{ }$\\
		$C^{\text{RR}}(i)$   & Total cost incurred by RR in Frame $i$\\
		\hline
	\end{tabular}
	\vspace*{10pt}
	\caption{Notation}
	\vspace*{-10pt}
	\label{table:proofs}
\end{table}

We use the following lemmas to prove Theorem \ref{thm:RR_adv}. 

The first lemma gives a lower bound on the number of requests that can be served by the edge server in the time-interval starting from a fetch to the subsequent eviction by OPT-OFF.
\begin{lemma}\label{lem:lemma_opt}
	
	If $r^*(n-1) = 0$, $r^*(t) = 1$ for $n \leq t \leq m$ and $r^*(m+1) = 0$, then,  
	$
	\displaystyle\sum_{l=n}^m \underbar{\text{$x$}}_l \geq M+	\displaystyle\sum_{l=n}^m c_l.
	$
\end{lemma}
\begin{IEEEproof}
	The cost incurred by OPT-OFF in $n \leq t \leq m$ is $M+\displaystyle\sum_{l=n}^m c_l+\displaystyle\sum_{l=n}^m \delta_l.$ We prove Lemma \ref{lem:lemma_opt} by contradiction. Let us assume that $\displaystyle\sum_{l=n}^m \underbar{\text{$x$}}_l < M+\displaystyle\sum_{l=n}^m c_l$. We construct another policy $\eta$ which behaves same as OPT-OFF except that $r_{\eta}(t) = 0$ for $n \leq t \leq m$.
	The total cost incurred by $\eta$ in $n \leq t \leq m$ is $\displaystyle\sum_{l=n}^m \underbar{\text{$x$}}_l+\displaystyle\sum_{l=n}^m\delta_l.$ It follows that
	$C^{\eta}(n,m)-C^{\text{OPT-OFF}}(n,m)= \displaystyle\sum_{l=n}^m \underbar{\text{$x$}}_l-M-\displaystyle\sum_{l=n}^m c_l,$ which is negative by our assumption. This contradicts the definition of the OPT-OFF policy, thus proving the result.
\end{IEEEproof}
The next lemma shows that if the number of requests that can be served by the edge server in a time-interval exceeds a certain value (which is a function of the length of that time-interval) and the service is not hosted  at the beginning of this time-interval, then OPT-OFF fetches the service at least once in the time-interval. 
\begin{lemma}\label{lem:OPT_download}
	If $r^*(n-1) = 0$, and $\displaystyle\sum_{l=n}^m \underbar{\text{$x$}}_l \geq M+\displaystyle\sum_{l=n}^m c_l$, then OPT-OFF fetches the service at least once in the interval from time-slots $n$ to $m$.
\end{lemma}
\begin{IEEEproof}
	We prove Lemma \ref{lem:OPT_download} by contradiction. We  construct another policy $\eta$ which behaves same as OPT-OFF except that $r_{\eta}(t) = 1$ for $n \leq t \leq m$. 
	The total cost incurred by $\eta$ in $n \leq t \leq m$ is $C^{\eta}(n,m)=M+\displaystyle\sum_{l=n}^m c_l+\displaystyle\sum_{l=n}^m \delta_l$. It follows that
	$C^{\eta}(n,m)-C^{\text{OPT-OFF}}(n,m)= M+\displaystyle\sum_{l=n}^m c_l-\displaystyle\sum_{l=n}^m \underbar{\text{$x$}}_l,$ which is negative. Hence there exists at least one policy $\eta$ which performs better than OPT-OFF. This contradicts the definition of the OPT-OFF policy, thus proving the result.
\end{IEEEproof}

The next lemma provides a lower bound on the duration for which OPT-OFF hosts  the service once it is fetched.
\begin{lemma}\label{lem:OPT_slots}
	Once OPT-OFF fetches the service, it is hosted  for at least $\frac{M}{\kappa-c_{\text{min}}}$ slots.
\end{lemma}
\begin{IEEEproof}
	Suppose OPT-OFF fetches the service at the end of the $(n-1)^{\text{th}}$ time-slot and evicts it at the end of time-slot $m > n$. From Lemma \ref{lem:lemma_opt}, $\displaystyle\sum_{l=n}^m \underbar{\text{$x$}}_l \geq M+\displaystyle\sum_{l=n}^m c_l$. Since $\displaystyle\sum_{l=n}^m \underbar{\text{$x$}}_l\leq (m-n+1)\kappa$ and $\displaystyle\sum_{l=n}^m c_l\geq (m-n+1)c_{\text{min}}$, $(m-n+1)\kappa \geq M+(m-n+1)c_{\text{min}}$, i.e, $(m-n+1) \geq \frac{M}{\kappa-c_{\text{min}}}$. This proves the result.
\end{IEEEproof}

Our next lemma characterizes a necessary condition for RR to fetch the service. 
\begin{lemma}\label{lem:RR_slots}
	If $r^{\text{RR}}(n) = 0$ and $r^{\text{RR}}(n+1) = 1$, then, by the definition of the RR policy, $\exists \tau < n$ such that $\displaystyle\sum_{l=n-\tau+1}^n \underbar{\text{$x$}}_l\geq \displaystyle\sum_{l=n-\tau+1}^n c_l +M.$
	Let $\tau_{\text{min}}=\displaystyle\min_\tau \left\{\tau|\displaystyle\sum_{l=n-\tau+1}^n \underbar{\text{$x$}}_l \geq \displaystyle\sum_{l=n-\tau+1}^n c_l +M\right\}$. Then,  
	$\tau_{\text{min}} \geq \frac{M}{\kappa-c_{\text{min}}}.$
\end{lemma}
\begin{IEEEproof}
	Since $\displaystyle\sum_{l=n-\tau+1}^n \underbar{\text{$x$}}_l	\leq \tau \kappa$, $\tau_{\text{min}} \kappa \geq \tau_{\text{min}} c_{\text{min}}+M$, and the result follows. 
\end{IEEEproof}

The next lemma gives an upper bound on the number of requests that can be served by the edge server (subject to its computation power constraints) in a time-interval such that RR does not host  the service during the time-interval and fetches it in the last time-slot of the time-interval.

\begin{lemma}\label{lem:max_requests}
	Let $r^{\text{RR}}(n-1)=1$, $r^{\text{RR}}(t)=0$ for $n \leq t \leq m$ and $r^{\text{RR}}(m+1)=1$. Then for any $n\leq n'< m$,  $\displaystyle\sum_{l=n'}^m \underbar{\text{$x$}}_l < \displaystyle\sum_{l=n'}^{m-1} c_l+M+\kappa.$   
\end{lemma}
\begin{IEEEproof}
	Given $r^{\text{RR}}(m)=0$ and $r^{\text{RR}}(m+1)=1$, then for any $n\leq n'< m$, $\displaystyle\sum_{l=n'}^{m-1} \underbar{\text{$x$}}_l <  \displaystyle\sum_{l=n'}^{m-1} c_l+M.$ 
	By definition,
	\begin{align*}
	\displaystyle\sum_{l=n'}^m \underbar{\text{$x$}}_l&= \left(\displaystyle\sum_{l=n'}^{m-1} \underbar{\text{$x$}}_l\right)+\underbar{\text{$x$}}_{m}< \displaystyle\sum_{l=n'}^{m-1} c_l+M+\kappa.
	\end{align*}
	Thus proving the result.	
\end{IEEEproof}

Consider the event where both RR and OPT-OFF have hosted  the service in a particular time-slot. The next lemma states that given this, OPT-OFF evicts the service before RR. 

\begin{lemma}\label{lem:eviction}
	If $r^{\text{RR}}(n)=1$, $r^*(t) = 1$ for $n \leq t \leq m$, and $r^*(m+1) = 0$. Then,  $r^{\text{RR}}(t)=1$ for $n+1 \leq t \leq m+1$.
\end{lemma}
\begin{IEEEproof}
	We prove this by contradiction. Let $\exists \widetilde{m}<m$ such that $r^{\text{RR}}(\widetilde{m}+1)=0$. Then, from Algorithm \ref{algo:RR}, there exists an integer $\tau>0$  such that  $\displaystyle\sum_{l=\widetilde{m}-\tau+1}^{\widetilde{m}} \underbar{\text{$x$}}_l< \displaystyle\sum_{l=\widetilde{m}-\tau+1}^{\widetilde{m}} c_l-M.$ 
	
	The cost incurred by OPT-OFF in the interval $\widetilde{m}-\tau+1$ to $\widetilde{m}$ is $\displaystyle\sum_{l=\widetilde{m}-\tau+1}^{\widetilde{m}} c_l+\displaystyle\sum_{l=\widetilde{m}-\tau+1}^{\widetilde{m}}\delta_l$. 
	
	Consider an alternative policy $\eta$ for which $r_{\eta}(t)=0$ for $\widetilde{m}-\tau+1 \leq t \leq \widetilde{m}$, $r_{\eta}(\widetilde{m}+1)=1$, and $r_{\eta}(t)=r^*(t)$ otherwise. It follows that
	$C^{\eta}-C^{\text{OPT-OFF}}= \displaystyle\sum_{l=\widetilde{m}-\tau+1}^{\widetilde{m}} \underbar{\text{$x$}}_l + M  - \displaystyle\sum_{l=\widetilde{m}-\tau+1}^{\widetilde{m}} c_l$ which is negative by our assumption. This contradicts the definition of the OPT-OFF policy, thus proving the result.	
\end{IEEEproof}

Consider the case where both RR and OPT-OFF have hosted  the service in a particular time-slot. From the previous lemma, we know that, OPT-OFF evicts the service before RR. The next lemma gives a lower bound on the number of requests that can be served by the edge server (subject to its computation power constraints) in the interval which starts when OPT-OFF evicts the service from the edge server  and ends when RR evicts the service from the edge server. 
\begin{lemma}\label{lem:min_requests}
	Let $r^*(n-1)= 1, \ r^*(n)=0, \ r^{\text{RR}}(t) = 1$ for $n-1 \leq t \leq m$ and  $r^{\text{RR}}(m+1)=0$.Then for any $n\leq n'< m$, $\displaystyle\sum_{l=n'}^m \underbar{\text{$x$}}_l \geq 	\displaystyle\sum_{l=n'}^{m-1} c_l-M.$    	
\end{lemma}
\begin{IEEEproof}
	Given $r^{\text{RR}}(m)=1$ and $r^{\text{RR}}(m+1)=0$, then  for any $n\leq n'< m$, $\displaystyle\sum_{l=n'}^{m-1} \underbar{\text{$x$}}_l > 	\displaystyle\sum_{l=n'}^{m-1} c_l-M$. 
	By definition,
	\begin{align*}
	\displaystyle\sum_{l=n'}^m \underbar{\text{$x$}}_l&= \left(\displaystyle\sum_{l=n'}^{m-1} \underbar{\text{$x$}}_l\right)+\underbar{\text{$x$}}_{m}> \displaystyle\sum_{l=n'}^{m-1} c_l-M+0.
	\end{align*}
	Thus proving the result. 	
\end{IEEEproof}
Our next result states that RR does not fetch the service in the interval between an eviction and the subsequent fetch by OPT-OFF.

\begin{lemma}\label{lem:RR_no_download} If $r^*(n-1) = 1, \ r^*(t) = 0$ for $n \leq t \leq m$, and $r^*(m+1)=1$, then RR does not fetch the service in time-slots $n, n+1, \cdots, m-1$. 
\end{lemma}

\begin{IEEEproof}
	We prove this by contradiction. Let RR fetch the service in time-slot $t$ where $n\leq t \leq m-1.$ 
	
	Then from Algorithm \ref{algo:RR}, there exists an integer $\tau>0$ such that $t-\tau\geq n$ and $\displaystyle\sum_{l=t-\tau+1}^t \underbar{\text{$x$}}_l\geq \displaystyle\sum_{l=t-\tau+1}^t c_l+M.$
	If this condition is true, by Lemma \ref{lem:OPT_download},  OPT-OFF would have fetched the service at least once in the interval $t-\tau+1$ and $t$ for all $n\leq t \leq m-1.$ Hence RR does not fetch the service between $n$ and $m-1.$
\end{IEEEproof}

The next lemma states that in the interval between a fetch and the subsequent eviction by OPT-OFF, RR hosts  the service for at least one time-slot. 

\begin{lemma}\label{lem:RR_one_download}
	If $r^*(n-1) = 0$, $r^*(t) = 1$ for $n \leq t \leq m$ and $r^*(m+1) = 0$, then,
	for some  $n < t \leq m$, $r^{\text{RR}}(t) = 1.$
\end{lemma}
\begin{IEEEproof}
	We prove this by contradiction. Let $r^{\text{RR}}(t)=0$ for all $n \leq t \leq m$. Then by the definition of the RR policy,  $\displaystyle\sum_{l=t-\tau+1}^t \underbar{\text{$x$}}_l < \displaystyle\sum_{l=t-\tau+1}^t c_l+M$ for any $\tau>0$ and $\tau \leq t-n+1$.
	If we choose $t=m$   then  $\displaystyle\sum_{l=n}^m \underbar{\text{$x$}}_l < \displaystyle\sum_{l=n}^m c_l+M$, which is false from Lemma \ref{lem:lemma_opt}. This contradicts our assumption.  
\end{IEEEproof}

If both RR and OPT-OFF host  the service in a particular time-slot, from Lemma \ref{lem:eviction}, we know that OPT-OFF evicts the service before RR. The next lemma states that RR evicts the service before the next time OPT-OFF fetches it. 
\begin{lemma}\label{lem:RR_one_evict}
	If $r^*(n-1) = 1$, $r^*(t) = 0$ for $n \leq t \leq m$, $r^*(m+1) = 1$, and $r^{\text{RR}}(n-1)=1$, then, RR evicts the service by the end of time-slot $m$ and $r^{\text{RR}}(m+1) = 0$.
\end{lemma}
\begin{IEEEproof}
	We prove this by contradiction. Assume that RR does not evict the service in any time slot $t$ for all  $n \leq t \leq m$. 
	Then from the definition of the RR policy, $\displaystyle\sum_{l=t-\tau+1}^{t} \underbar{\text{$x$}}_l+M \geq \displaystyle\sum_{l=t-\tau+1}^{t} c_l$ for all $\tau$ such that $0 < \tau \leq t-n+1$.
	As a result, at $t=m$, $\displaystyle\sum_{l=n}^m \underbar{\text{$x$}}_l +M > \displaystyle\sum_{l=n}^m c_l$. 
	
	Given this, it follows that OPT-OFF will not evict the service at the end of time-slot $n-1$. 
	This contradicts our assumption.
	By Lemma \ref{lem:RR_no_download}, RR does not fetch the service in the interval between an eviction and the subsequent fetch by OPT-OFF. Therefore, $r^{\text{RR}}(m+1) = 0$.
\end{IEEEproof}

To compare the costs incurred by RR and OPT-OFF we divide time into frames $[1,t_1-1]$, $[t_1,t_2-1], [t_2,t_3-1],\cdots,$ where $t_i-1$ is the time-slot in which OPT-OFF fetches the service for the $i^{\text{th}}$ time for $i\in\{1,2,\cdots\}.$ Our next result characterizes the sequence of events that occur in any such frame.

\begin{figure}[ht]
	\centering
	\begin{tikzpicture}
	\foreach \x in {0,0.3,0.6,...,7.2}{
		\draw[] (\x,0) --  (\x+0.3,0);
		\draw[gray] (\x,-1mm) -- (\x,1mm);
	}
	
	\draw[gray] (7.2,-1mm) -- (7.2,1mm);
	\draw[-latex] (0.25,10mm) -- node[above=5mm]{$t_{i}-1$} (0.25,0mm);
	\draw[-latex] (6.85,10mm) -- node[above=5mm]{$t_{i+1}-1$} (6.85,0mm);
	\draw[-latex] (4.15,0mm) -- node[above=5mm]{$\tau$} (4.15,10mm);
	\draw[-latex,color=red] (2.35,10mm) -- node[above=5mm]{$t_f^{\text{RR}}$} (2.35,0mm);
	\draw[-latex,color=red] (5.65,0mm) -- node[above=5mm]{$t_e^{\text{RR}}$} (5.65,10mm);
	
	\draw[gray] (7.2,-1mm) -- (7.2,1mm);
	\draw[-latex] (0.25,10mm) -- node[above=5mm]{} (0.25,0mm);
	\draw[-latex] (6.85,10mm) -- node[above=5mm]{} (6.85,0mm);
	\draw[-latex] (4.15,0mm) -- node[above=5mm]{} (4.15,10mm);
	\draw[-latex,color=red] (2.35,10mm) -- node[above=5mm]{} (2.35,0mm);
	\draw[-latex,color=red] (5.65,0mm) -- node[above=5mm]{} (5.65,10mm);
	
	\draw[<->,color=black] (0.3,-2mm) -- node[below=0mm,pos=0.5]{Frame $i$} (6.9,-2mm);
	
	\draw[black] (0.3,-1mm) -- (0.3,-3mm);
	\draw[black] (6.9,-1mm) -- (6.9,-3mm);
	
	\node[] at (7.7,-0.9) {OPT-OFF};
	
	\filldraw[fill=black!] (0.25,-1) rectangle (4.15,-0.7);
	\filldraw[fill=white!40] (4.15,-1) rectangle (6.85,-0.7);
	\node[] at (7.25,-1.4) {RR};
	\filldraw[fill=white!40] (0.25,-1.5) rectangle (2.4,-1.2);
	\filldraw[fill=red!] (2.4,-1.5) rectangle (5.7,-1.2);
	\filldraw[fill=white!40] (5.7,-1.5) rectangle (6.85,-1.2);

	\end{tikzpicture}
	\caption{Illustration of Lemma \ref{lem:frameStructure} showing download and eviction by OPT-OFF and  RR in the $i^{\text{th}}$ frame. Downward arrows represent fetches, upward arrows indicate evictions. Black and red arrows correspond to the OPT-OFF and RR policy respectively. The two bars below the timeline indicate the state of the edge server  under OPT-OFF and RR. The solid black and solid red portions represent the intervals during with OPT-OFF and RR host  the service on the edge server respectively}
	\label{fig:OPT_RR_frame}
\end{figure}
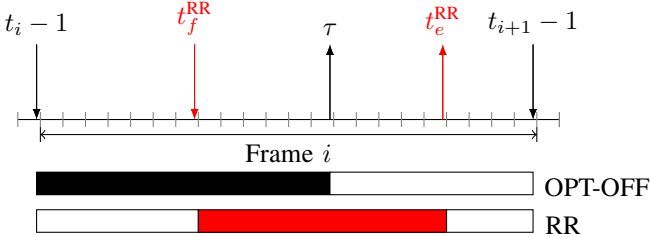

\begin{lemma}
	\label{lem:frameStructure}
	Consider the interval $[t_i, t_{i+1}-1]$ such that OPT-OFF fetches the service at the end of time-slot $t_{i}-1$ and fetches it again the end of time-slot $t_{i+1}-1$. By definition, there exists $\tau \in [t_i, t_{i+1}-2]$ such that OPT-OFF evicts the service in time-slot $\tau$. 
	RR fetches and evicts the service exactly once each in $[t_1, t_2-1]$. The fetch by RR is in time-slot $t_{f}^{\text{RR}} $ such that $t_1 \leq t_{f}^{\text{RR}} \leq \tau$ and
	the eviction by RR is in time-slot $t_{e}^{\text{RR}}$ such that $\tau < t_{e}^{\text{RR}} < t_2$ (Figure \ref{fig:OPT_RR_frame}).
\end{lemma}

\begin{IEEEproof}
	Without loss of generality, we prove the result for $i=1$.
	Since $r^*(t_1-1) = 0$, $r^*(t) = 1$ for $t_1 \leq t \leq \tau$ and $r^*( \tau) = 0$ then by Lemma \ref{lem:RR_one_download}, $r^{\text{RR}}(t_{f}^{\text{RR}}) = 1$ for some  $t_1 < t_{f}^{\text{RR}} \leq \tau.$ In addition, by Lemma \ref{lem:RR_one_evict}, $r^{\text{RR}}(t_1) = 0$. Therefore, RR fetches the service at least once in the interval $[t_1, t_2-1]$.
	
	By Lemma \ref{lem:eviction}, if $t_{f}^{\text{RR}} < \tau$, since both RR and OPT-OFF host  the service during time-slot $t_{f}^{\text{RR}}+1$, OPT-OFF evicts the service before RR, therefore, once fetched, RR does not evict the service before time-slot $\tau+1$, i.e., $r^{\text{RR}}(t)=1$ for $t_{f}^{\text{RR}}+1 \leq t \leq \tau+1.$

	Since $r^*(\tau) = 1$, $r^*(t) = 0$ for $\tau+1 \leq t \leq t_2-1$ and $r^*(t_2) = 1$, then by Lemma \ref{lem:RR_one_evict}, RR evicts the service in time-slot $t_{e}^{\text{RR}}$ such that $\tau < t_{e}^{\text{RR}} \leq t_2-1$.
	
	In addition, once evicted at $t_{e}^{\text{RR}} \leq t_2-1$, RR does not fetch it again in the before time-slot $t_2$ by Lemma \ref{lem:RR_no_download}. 
	
	This completes the proof.
\end{IEEEproof}

\begin{IEEEproof}[Proof of Theorem \ref{thm:RR_adv} ] As mentioned above, to compare the costs incurred by RR and OPT-OFF we divide times into frames $[1,t_1-1]$, $[t_1,t_2-1], [t_2,t_3-1],\ldots,$ where $t_i-1$ is the time-slot in which OPT-OFF downloads the service for the $i^{\text{th}}$ time for $i\in\{1,2,\ldots, k\}.$

	For convenience, we account for the fetch cost incurred by OPT-OFF in time-slot $t_i$ in the cost incurred by OPT-OFF in Frame $i$. Given this, the cost under RR and OPT-OFF is the same for $[1,t_1-1]$ (Frame 0) since both policies don't host  the service in this period. 
	
	Note that if the total number of fetches made by OPT-OFF is $k<\infty$, there are exactly $k+1$ frames (including Frame 0). The $(k+1)^{\text{th}}$ frame either has no eviction by OPT-OFF or OPT-OFF evicts and then never fetches the service. 
	
	We now focus on Frame $i$, such that $0<i<k$, where $k$ is the total number of fetches made by OPT-OFF.

	Without loss of generality, we focus on Frame 1. Recall the definitions of $\tau$, $t_e^{\text{RR}}$, and $t_f^{\text{RR}}$ from Lemma \ref{lem:frameStructure}, also seen in Figure \ref{fig:OPT_RR_frame}. By Lemma \ref{lem:frameStructure}, we have that RR fetches and evicts the service exactly once each in $[t_1, t_2-1]$ such that the fetch by RR is in time-slot $t_{f}^{\text{RR}} $ such that $t_1 \leq t_{f}^{\text{RR}} \leq \tau$ and 
	the eviction by RR is in time-slot $t_{e}^{\text{RR}}$ such that $\tau < t_{e}^{\text{RR}} < t_2$.
	
	Both OPT-OFF and RR makes one fetch in the frame. Hence the difference in the fetch costs is zero. We now focus on the service and rent cost incurred by the two policies. 
	
	By Lemma \ref{lem:max_requests}, the number of requests that can be served by the edge server in $[t_1, t_{f}^{\text{RR}}]$ is at most $M+\displaystyle \sum_{l = t_1}^{t_{f}^{\text{RR}}} c_l +\kappa.$ 
	Since RR does not host  the service in $[t_1, t_{f}^{\text{RR}}]$, the rent cost incurred in $[t_1, t_{f}^{\text{RR}}]$ by RR is zero and the service cost incurred in $[t_1, t_{f}^{\text{RR}}]$ by RR 
	is at most $M+\displaystyle \sum_{l = t_1}^{t_{f}^{\text{RR}}-1} c_l +\kappa + \displaystyle \sum_{l = t_1}^{t_{f}^{\text{RR}}} \delta_l$. OPT-OFF rents storage in $[t_1, t_{f}^{\text{RR}}]$ at cost $ \displaystyle \sum_{l = t_1}^{t_{f}^{\text{RR}}} c_l$ and incurs a service cost of $\displaystyle \sum_{l = t_1}^{t_{f}^{\text{RR}}} \delta_l$ in $[t_1, t_{f}^{\text{RR}}]$. 
	Hence difference in the service and rent cost incurred by RR and OPT-OFF in $[t_1, t_{f}^{\text{RR}}]$   is at most $M+\kappa-c_{\text{min}}.$
	
	The service and rent cost incurred by OPT-OFF and RR in $[t_{f}^{\text{RR}}+1, \tau]$ are equal. 
	
	By Lemma \ref{lem:min_requests}, the number of requests that can be served by the edge server in $[\tau+1, t_{e}^{\text{RR}}]$ is at least $ \displaystyle \sum_{l = \tau+1}^{t_{e}^{\text{RR}}-1} c_l-M.$ The service cost incurred by OPT-OFF in $[\tau+1, t_{e}^{\text{RR}}]$ is at least $\displaystyle \sum_{l = \tau+1}^{t_{e}^{\text{RR}}-1} c_l-M + \displaystyle \sum_{l = \tau+1}^{t_{e}^{\text{RR}}} \delta_l$ and the rent cost incurred by OPT-OFF in $[\tau+1, t_{e}^{\text{RR}}]$ is zero. The rent cost incurred by RR in $[\tau+1, t_{e}^{\text{RR}}]$ is $\displaystyle \sum_{l = \tau+1}^{t_{e}^{\text{RR}}} c_l$ and the service cost incurred by RR in $[\tau+1, t_{e}^{\text{RR}}]$ is $\displaystyle \sum_{l = \tau+1}^{t_{e}^{\text{RR}}} \delta_l$. Hence difference in the service and rent cost incurred by RR and OPT-OFF in $[t_1, t_{f}^{\text{RR}}]$  is at most $M+c_{\text{max}}.$
	
	The service and rent cost incurred by OPT-OFF and RR in $[t_{e}^{\text{RR}}+1, t_2-1]$ are equal. 
	
	Let $C^{\text{RR}}(i),$ $C^{\text{OPT-OFF}}(i)$ denote the costs incurred in the $i^{\text{th}}$ frame by RR and OPT-OFF respectively.
	We therefore have that,
	\begin{align*}
	C^{\text{RR}}(i)-C^{\text{OPT-OFF}}(i)&\leq 2M+\kappa+c_{\text{max}}-c_{\text{min}}.
	\end{align*}

	Since $c_{\text{max}}\leq M+\kappa$, we have that,
	\begin{align}
	C^{\text{RR}}(i)-C^{\text{OPT-OFF}}(i)&\leq 3M+2\kappa.
	\label{ineq:bound1_RR}
	\end{align}
	
	By Lemma \ref{lem:OPT_slots}, once OPT-OFF downloads the service, it is hosted  for at least $\frac{M}{\kappa-c_{\text{min}}}$ slots. Therefore,
	\begin{align}
	&C^{\text{OPT-OFF}}(i)\geq M+\frac{Mc_{\text{min}}}{\kappa-c_{\text{min}}} \implies C^{\text{OPT-OFF}}(i)\geq \frac{M\kappa}{\kappa-c_{\text{min}}}\nonumber\\
	&M\leq \frac{\kappa-c_{\text{min}}}{\kappa}C^{\text{OPT-OFF}}(i) \text{ and } \kappa\leq \frac{\kappa-c_{\text{min}}}{M}C^{\text{OPT-OFF}}(i).
	\label{ineq:bound1_OPT}
	\end{align}
	
	From \eqref{ineq:bound1_RR} and \eqref{ineq:bound1_OPT}, 
	\begin{align}
	C^{\text{RR}}(i) &\leq \left(4-\frac{3c_{\text{min}}}{\kappa}\right)C^{\text{OPT-OFF}}(i)+2\left(\frac{\kappa-c_{\text{min}}}{M}\right)C^{\text{OPT-OFF}}(i)\nonumber\\
	& \leq \left(4+\frac{2(\kappa-c_{\text{min}})}{M}-\frac{3c_{\text{min}}}{\kappa}\right)C^{\text{OPT-OFF}}(i).
	\end{align}

	This completes the characterization for Frame $1$ to $k-1$.
	
	We now focus on Frame $k$, which is the last frame. There are two possible cases, one where OPT-OFF evicts the service in Frame $k$, in which case the analysis for Frame $k$ is identical to that of Frame $1$, and the other when OPT-OFF does not evict the service in Frame $k$. We now focus on the latter.

	Given that OPT-OFF downloads the service in time-slot $t_{k}-1$, there exists $m>t_{k}$ such that
	$\displaystyle\sum_{l=t_k}^m \underbar{\text{$x$}}_l \geq M+\displaystyle\sum_{l=t_k}^m c_l$.
	By Step~8 in Algorithm \ref{algo:RR}, RR downloads the service at the end of
	time-slot $m$.
	Let $\tau_k = m-t_k$. By Lemma \ref{lem:max_requests}, the number of requests that can be served by the edge server during these $\tau_k$ time-slots is at most $M+\displaystyle\sum_{l=t_k}^{m-1} c_l+\kappa.$
	Since RR does not host  the service during these  $\tau_k$ time-slots, the rent cost incurred by RR is zero and the service cost incurred by RR is at most $M+\displaystyle\sum_{l=t_k}^{m-1} c_l+\kappa + \displaystyle \sum_{l = t_k}^{m} \delta_l$. OPT-OFF rents storage during these $\tau_k$ time-slots at cost $\displaystyle \sum_{l = t_k}^{m} c_l$ and the service cost incurred by OPT-OFF is $\displaystyle \sum_{l = t_k}^{m - 1} \delta_l$. There is no difference between the cost of RR and OPT-OFF after the first $\tau_k$ slots in Frame $k$. It follows that
	\begin{align}
	C^{\text{RR}}(k)-C^{\text{OPT-OFF}}(k)\leq M+\kappa-c_{\text{min}}.
	\label{ineq:bound1_RRk}
	\end{align}
	From \eqref{ineq:bound1_OPT} and \eqref{ineq:bound1_RRk},

	\begin{align}
	C^{\text{RR}}(k) &\leq \left(2-\frac{c_{\text{min}}}{\kappa}\right)C^{\text{OPT-OFF}}(k)+\left(\frac{\kappa-c_{\text{min}}}{M}\right)C^{\text{OPT-OFF}}(k)\nonumber\\
	 &< \left(4+\frac{2(\kappa-c_{\text{min}})}{M}-\frac{3c_{\text{min}}}{\kappa}\right)C^{\text{OPT-OFF}}(k).
	\end{align}

	Stitching together the results obtained for all frames, the result follows.  
	
\end{IEEEproof}

\subsection{Proof of Theorem \ref{thm:any_online}}
\begin{IEEEproof}
	Let $\mathcal{P}$ be a given deterministic online policy and $C^\mathcal{P}(a,b)$ be the cost incurred by this policy for the request sequence $a$ and rent cost sequence $b$.
	
	We first focus on the case where $\mathcal{P}$ does not host  the service during the first time-slot. 
	
	We define $t_\text{fetch}^{(1)}\geq1$ as the first time the policy $\mathcal{P}$ fetches the service when there are $\kappa$ request arrivals each in the first $t_\text{fetch}^{(1)}$ time-slots. Since $\mathcal{P}$ is an online deterministic policy, the value of $t_\text{fetch}^{(1)}$ can be computed a-priori.
	
	Consider the arrival process $a$ with $\kappa$ request arrivals each in the first $t_\text{fetch}^{(1)}$ time-slots and no arrivals thereafter.
	Consider a renting cost sequence $b$ with  $c_{\text{min}}$ in the first $t_\text{fetch}^{(1)}$ time-slots.
	It follows that $C^{\mathcal{P}}(a,b) \geq \kappa t_\text{fetch}^{(1)} + M + c_{\text{min}}.$
	
	Consider an alternative policy ALT which hosts  the service in time-slots $1$ to $t_\text{fetch}^{(1)}$ and does not host  it thereafter. It follows that $C^{\text{ALT}}(a,b) = c_{\text{min}} t_\text{fetch}^{(1)} + M.$
	By definition,
	$
	\rho^{\mathcal{P}} \geq \dfrac{\kappa t_\text{fetch}^{(1)} + M + c_{\text{min}}}{c_{\text{min}} t_\text{fetch}^{(1)} + M}. 
	$
	Therefore,
	\begin{align*}
	\rho^{\mathcal{P}} \geq
	\begin{cases*}
	1 + \dfrac{\kappa}{c_{\text{min}} + M} \hspace{1em} \text{ if } \kappa \geq \dfrac{c_{\text{min}}(c_{\text{min}}+M)}{M} \\
	\dfrac{\kappa}{c_{\text{min}}} \hspace{4.8em} \text{ otherwise. }
	\end{cases*}
	\end{align*}
	
	Next, we focus on the case where $\mathcal{P}$ hosts  the service during the first time-slot. We define $t_\text{evict}^{(1)}\geq1$ as the first time the policy $\mathcal{P}$ evicts the service when there are no request arrivals each in the first $t_\text{evict}^{(1)}$ time-slots. Since $\mathcal{P}$ is an online deterministic policy, the value of $t_\text{evict}^{(1)}$ can be computed a-priori.
	
	Consider the arrival process $a$ with no request arrivals each in the first $t_\text{evict}^{(1)}$ time-slots and $\kappa$ arrivals in time-slot $t_\text{evict}^{(1)}+1$.
	Consider the renting cost sequence $b$ with $c_{\text{max}}$ each in the first $t_\text{evict}^{(1)}$ time-slots and $c_{\text{min}}$  in time-slot $t_\text{evict}^{(1)}+1$.
	It follows that $C^{\mathcal{P}}(a,b) \geq c_{\text{max}} t_\text{evict}^{(1)} + M + \kappa.$ Consider an alternative policy ALT which does not host  the service in time-slots $1$ to $t_\text{evict}^{(1)}$, hosts  it in time-slot $t_\text{evict}^{(1)}+1$ and does not host  it thereafter. It follows that $C^{\text{ALT}}(a,b) = M + c_{\text{min}}.$
	By definition,
	\begin{align*}
	\rho^{\mathcal{P}} \geq \dfrac{c_{\text{max}} t_\text{evict}^{(1)} + M + \kappa}{M+c_{\text{min}}} 
	&\geq 1 + \dfrac{\kappa+c_{\text{max}}-c_{\text{min}}}{c_{\text{min}}+M}.
	\end{align*}
\end{IEEEproof}


\subsection{Proof of Theorem \ref{thm:RR_stochastic_theorem}}
We use the following lemmas to prove Theorem \ref{thm:RR_stochastic_theorem}. 

\begin{lemma}
	\label{lemma:optimal_causal}
	Let $X_t$ be the number  of requests arriving in time-slot $t$, $\nu = \mathbb{E}[X_t]$,  $\underbar{$X$}_t=\min\{X_t,\kappa\}$ and $\mu =  \mathbb{E}[\underbar{$X$}_t]$. 
	Let $R_t$ be the rent cost per time-slot, $\{R_t\}_{t\geq 1}$ is the sequence of negatively associated random variables and $c= \mathbb{E}[R_t]$.
	Under Assumption \ref{assum_stochastic}, let $\mathbb{E}[ C_t^{\text{OPT-ON}}]$ be the cost per time-slot incurred by the OPT-ON policy. Then,
	\begin{align*}
	\mathbb{E}[ C_t^{\text{OPT-ON}}] \geq \min\{c + \nu - \mu, \nu\}.
	\end{align*}
\end{lemma}
\begin{IEEEproof}
	If the service is hosted  at the edge in time-slot $t$, the expected cost incurred is at least $\mathbb{E}[X_t +R_t- \min\{X(t) , \kappa\}]$ = $c + \nu - \mu$. 
	
	If the service is not hosted  at the edge server, the expected cost incurred is at least $\nu$. This proves the result.
\end{IEEEproof}

\begin{lemma}\label{lem:Hoeffding}
	Let $X_t$ be the number  of requests arriving in time-slot $t$, $\underbar{$X$}_t=\min\{X_t,\kappa\}$ and $\mu =  \mathbb{E}[\underbar{$X$}_t]$. 
	Let $R_t$ be the rent cost per time-slot, $\{R_t\}_{t\geq 1}$ is the sequence of negatively associated random variables and $c= \mathbb{E}[R_t]$. Define $Y_l=\underbar{$X$}_l-R_l$, and  $Y=\sum\limits_{l=t-\tau+1}^t Y_l$ then $Y$ satisfies,
	
	\noindent for $(c-\mu) \tau + M > 0$,
	\begin{align*}
	\mathbb{P}\left(Y \geq M\right)  &\leq \exp\left(-2\frac{((c-\mu)\tau+M)^2}{\tau (\kappa+c_{\text{max}}-c_{\text{min}})^2}\right), 
	\end{align*}
	and for $(\mu-c) \tau + M > 0$,
	\begin{align*}
	\mathbb{P}\left(Y \leq \tau c-M\right)  &\leq \exp\left(-2\frac{((\mu-c)\tau+M)^2}{\tau (\kappa+c_{\text{max}}-c_{\text{min}})^2}\right). 
	\end{align*}
\end{lemma}
\begin{IEEEproof}   Using i.i.d. condition of $\{X_t\}_{t\geq 1}$  and  negatively associativity of $\{R_t\}_{t\geq 1}$, it follows that for $s>0$, $\mathbb{E}[\exp(sY)]\leq \prod\limits_{l=t-\tau+1}^t \mathbb{E}[\exp(sY_l)]$.  Moreover,	 $Y_l\in [ -c_{\text{max}}, \kappa-c_{\text{min}}]$. Then the result  follows by Hoeffding's inequality  \cite{hoeffding1994probability, wajc2017negative}. 
\end{IEEEproof}
\vspace{-1pt}
\begin{IEEEproof}[Proof of Lemma \ref{lemma:difference_RRstochastic}] We first consider the case when $\mu>c$. We define the following events

$E_{t_1,t_2}: \displaystyle\sum_{l=t_1}^{t_2} \underbar{$X$}_l \leq \displaystyle\sum_{l=t_1}^{t_2} R_l - M$, $E^{\tau} =\displaystyle\bigcup_{t_1=1}^{\tau} E_{t_1,\tau}$, $E =\displaystyle\bigcup_{\tau=t-\lceil\frac{\lambda M}{\mu-c}\rceil}^{t-1} E^{\tau}$, $F: \displaystyle\sum_{l=t-\lceil\frac{\lambda M}{\mu-c}\rceil}^{t-1} \underbar{$X$}_l \geq \displaystyle\sum_{l=t-\lceil\frac{\lambda M}{\mu-c}\rceil}^{t-1} R_l+M$.  \\
	
	By Lemma \ref{lem:Hoeffding}, it follows that 
	$\mathbb{P}(E_{t_1,t_2})\leq \exp\left(-2\frac{(\mu-c)^2 (t_2-t_1+1)}{(\kappa+c_{\text{max}}-c_{\text{min}})^2}\right),$ 
	and therefore,
	\begin{align}
	\mathbb{P}(E^{\tau})&\leq \displaystyle \sum_{t_1=1}^{\tau-\lceil\frac{M}{c}\rceil+1} \exp\left(-2\frac{(\mu-c)^2 (\tau-t_1+1)}{ (\kappa+c_{\text{max}}-c_{\text{min}})^2}\right)\nonumber\\
	& \leq \frac{\exp\left(-2\frac{(\mu-c)^2 \frac{M}{c}}{ (\kappa+c_{\text{max}}-c_{\text{min}})^2}\right)}{1-\exp\left(-2\frac{(\mu-c)^2}{ (\kappa+c_{\text{max}}-c_{\text{min}})^2}\right)}. \label{eq:E_tau}
	\end{align}
	Using \eqref{eq:E_tau} and the union bound,
	
	\begin{align}
	\mathbb{P}(E)\leq \bigg\lceil\frac{\lambda M}{\mu-c}\bigg\rceil \frac{\exp\left(-2\frac{(\mu-c)^2 \frac{M}{c}}{ (\kappa+c_{\text{max}}-c_{\text{min}})^2}\right)}{1-\exp\left(-2\frac{(\mu-c)^2}{ (\kappa+c_{\text{max}}-c_{\text{min}})^2}\right)}. \label{eq:E}
	\end{align}
	By Lemma \ref{lem:Hoeffding}, 
	\begin{align}
	\mathbb{P}(F^c)&\leq \exp\left(-2\frac{((\mu-c)\lceil\frac{\lambda M}{\mu-c}\rceil-M)^2}{\frac{\lambda M}{\mu-c}  (\kappa+c_{\text{max}}-c_{\text{min}})^2}\right)\nonumber\\
	&  \leq \exp\left(-2\frac{(\lambda-1)^2M(\mu-c)}{\lambda  (\kappa+c_{\text{max}}-c_{\text{min}})^2}\right). \label{eq:F}
	\end{align}
	By \eqref{eq:E} and \eqref{eq:F}, 
	\begin{align}
	\mathbb{P}(E^c \cap F)\geq 1 &- \bigg\lceil\frac{\lambda M}{\mu-c}\bigg\rceil \frac{\exp\left(-2\frac{(\mu-c)^2 \frac{M}{c}}{ (\kappa+c_{\text{max}}-c_{\text{min}})^2}\right)}{1-\exp\left(-2\frac{(\mu-c)^2}{ (\kappa+c_{\text{max}}-c_{\text{min}})^2}\right)} -\nonumber\\
	& \exp\left(-2\frac{(\lambda-1)^2M(\mu-c)}{\lambda  (\kappa+c_{\text{max}}-c_{\text{min}})^2}\right). \label{eq:EcapF}
	\end{align}
	Consider the event $G = E^c \cap F$ and the following three cases. 
	
	Case 1: The service is hosted  at the edge during time-slot $t -\big\lceil\frac{\lambda M}{\mu-c}\big\rceil$: Conditioned on $E^c$, by the properties of the RR  policy, the service is not evicted from the edge server  in time-slots $t - \big\lceil\frac{\lambda M}{\mu-c}\big\rceil$ to $t-1$. It follows that in this case, the service is hosted at the edge server  during time-slot $t$.
	
	Case 2: The service is not hosted  at the edge during time-slot $t - \big\lceil\frac{\lambda M}{\mu-c}\big\rceil$ and is fetched in time-slot $\tilde{\tau}$ such that $t - \big\lceil\frac{\lambda M}{\mu-c}\big\rceil + 1 \leq \tilde{\tau} \leq t-2$: Conditioned on $E^c$, by the properties of the RR policy, the service is not evicted from the edge server  in time-slots $\tilde{\tau}+1$ to $t-1$. It follows that in this case, the service is hosted at the edge during time-slot $t$.
	
	Case 3: The service is not hosted  at the edge during time-slot $t - \big\lceil\frac{\lambda M}{\mu-c}\big\rceil$ and is not fetched in time-slots $t - \big\lceil\frac{\lambda M}{\mu-c}\big\rceil + 1$ to $t-2$: In this case, in time-slot $t-1$, $t_{\text{evict}} \leq t - \big\lceil\frac{\lambda M}{\mu-c}\big\rceil$. Conditioned on $F$, by the properties of the RR policy, condition in Step 8 in Algorithm \ref{algo:RR}   is satisfied for $\tau = t - \big\lceil\frac{\lambda M}{\mu-c}\big\rceil$.  It follows that in this case, the service is fetched in time-slot $t-1$ and therefore, the service is  hosted at the edge during time-slot $t$.
	
	We thus conclude that conditioned on $G = E^c \cap F$, the service is hosted  at the edge during time-slot $t$. 
	We now compute the expected cost incurred by the RR policy. By definition,
	\begin{align*}
	\mathbb{E}[C_t^{\text{RR}}] 
	=  \mathbb{E}[C_t^{\text{RR}}|G] \mathbb{P}(G) +\mathbb{E}[C_t^{\text{RR}}|G^c] \times \mathbb{P}(G^c).
	\end{align*}
	Note that, 
	$
	\mathbb{E}[C_t^{\text{RR}}|G] = c + \nu-\mu, \ \mathbb{E}[C_t^{\text{RR}}|G^c] \leq M + c + \nu.
	$
	Therefore,
	\begin{align}
	\mathbb{E}[C_t^{\text{RR}}] 
	= &  c + \nu-\mu + (M+\mu) \mathbb{P}(G^c) \nonumber \\
	\leq &  c + \nu-\mu +\nonumber\\
	& (M+\mu) \times  \bigg(\bigg\lceil\frac{\lambda M}{\mu-c}\bigg\rceil\frac{\exp\left(-2\frac{(\mu-c)^2\frac{M}{c}}{\kappa^2}\right)}{1-\exp\left(-2\frac{(\mu-c)^2}{ (\kappa+c_{\text{max}}-c_{\text{min}})^2}\right)}\nonumber\\
	&\hspace{1em}+\exp\left(-2\frac{(\lambda-1)^2M(\mu-c)}{\lambda  (\kappa+c_{\text{max}}-c_{\text{min}})^2}\right)\bigg). \label{eq:finalBound}
	\end{align}
	We optimize over $\lambda>1$ to get the tightest possible bound. By Lemma \ref{lemma:optimal_causal} and \eqref{eq:finalBound}, we have the result for RR.

	Next, we consider the case when $\mu < c$. We define the following events
	
	$F_{t_1,t_2}: \displaystyle\sum_{l=t_1}^{t_2} \underbar{$X$}_l \geq \displaystyle\sum_{l=t_1}^{t_2} R_l+M$, $F^{\tau}=\displaystyle\bigcup_{t_1=1}^{\tau} F_{t_1,\tau}$, $F_{t-1}=\displaystyle\bigcup_{\tau=t-\lceil\frac{\lambda M}{c-\mu}\rceil}^{t-1} F^{\tau}$, $F_{t}=\displaystyle\bigcup_{\tau=t-\lceil\frac{\lambda M}{c-\mu}\rceil+1}^{t} F^{\tau}$,                             $E: \displaystyle\sum_{l=t-\frac{\lambda M}{c-\mu}}^{t-1} \underbar{$X$}_l+M < \displaystyle\sum_{l=t-\frac{\lambda M}{c-\mu}}^{t-1} R_l$.\\
	
	By Lemma \ref{lem:Hoeffding}, it follows that  $\mathbb{P}(F_{t_1,t_2})\leq \exp\left(-2\frac{(c-\mu)^2 (t_2-t_1+1)}{ (\kappa+c_{\text{max}}-c_{\text{min}})^2}\right),$ and therefore, 
	\begin{align}
	\mathbb{P}(F^{\tau})&\leq \displaystyle \sum_{t_1=1}^{\tau-\lceil\frac{M}{\kappa-c}\rceil+1} \exp\left(-2\frac{(c-\mu)^2 (\tau-t_1+1)}{ (\kappa+c_{\text{max}}-c_{\text{min}})^2}\right)\nonumber\\
	&\leq \frac{\exp\left(-2\frac{(c-\mu)^2 \frac{M}{\kappa-c}}{\kappa^2}\right)}{1-\exp\left(-2\frac{(c-\mu)^2}{ (\kappa+c_{\text{max}}-c_{\text{min}})^2}\right)}. \label{eq:F_tau}
	\end{align}
	Using \eqref{eq:F_tau} and the union bound, $\mathbb{P}(F_{t})$ and $\mathbb{P}(F_{t-1})$ are upper bounded by
	\begin{equation}
	\bigg\lceil\frac{\lambda M}{c-\mu}\bigg\rceil \frac{\exp\left(-2\frac{(c-\mu)^2 \frac{M}{\kappa-c}}{\kappa^2}\right)}{1-\exp\left(-2\frac{(c-\mu)^2}{ (\kappa+c_{\text{max}}-c_{\text{min}})^2}\right)}. \label{eq:F2}
	\end{equation}
	By Lemma \ref{lem:Hoeffding},  
	\begin{align}
	\mathbb{P}(E^c)&\leq \exp\left(-2\frac{((c-\mu)\lceil\frac{\lambda M}{c-\mu}\rceil-M)^2}{\frac{\lambda M}{c-\mu}  (\kappa+c_{\text{max}}-c_{\text{min}})^2}\right)\nonumber\\
	& \leq \exp\left(-2\frac{(\lambda-1)^2(c-\mu)M}{\lambda  (\kappa+c_{\text{max}}-c_{\text{min}})^2}\right). \label{eq:E2}
	\end{align} 
	By \eqref{eq:F2} and \eqref{eq:E2},
	\begin{align}
	\mathbb{P}(F^c_t \cap F^c_{t-1} &\cap E)\geq 1- \exp\left(-2\frac{(\lambda-1)^2(c-\mu)M}{\lambda  (\kappa+c_{\text{max}}-c_{\text{min}})^2}\right)\nonumber\\
	&-2\bigg\lceil\frac{\lambda M}{c-\mu}\bigg\rceil \frac{\exp\left(-2\frac{(c-\mu)^2 \frac{M}{\kappa-c}}{ (\kappa+c_{\text{max}}-c_{\text{min}})^2}\right)}{1-\exp\left(-2\frac{(c-\mu)^2}{ (\kappa+c_{\text{max}}-c_{\text{min}})^2}\right)}.
    \label{eq:FcapE2}
	\end{align}
	Consider the event $G = F^c_t \cap F^c_{t-1} \cap E$ and the following three cases. 
	
	Case 1: The service is not hosted  at the edge during time-slot $t -\big\lceil\frac{\lambda M}{c-\mu}\big\rceil$: Conditioned on $F^c$, by the properties of the RR  policy, the service is not fetched in time-slots $t - \big\lceil\frac{\lambda M}{c-\mu}\big\rceil$ to $t-1$. It follows that in this case, the service is not  hosted  at the edge during time-slot $t$.
	
	Case 2: The service is hosted  at the edge during time-slot $t - \big\lceil\frac{\lambda M}{c-\mu}\big\rceil$ and is evicted in time-slot $\tilde{\tau}$ such that $t - \big\lceil\frac{\lambda M}{c-\mu}\big\rceil + 1 \leq \tilde{\tau} \leq t-2$: Conditioned on $F^c$, by the properties of the RR policy, the service is not fetched in time-slots $\tilde{\tau}+1$ to $t-1$. It follows that in this case, the service is not  hosted  at the edge during time-slot $t$.
	
	
	Case 3: The service is hosted  at the edge during time-slot $t - \big\lceil\frac{\lambda M}{c-\mu}\big\rceil$ and is not evicted in time-slots $t - \big\lceil\frac{\lambda M}{c-\mu}\big\rceil + 1$ to $t-2$: In this case, in time-slot $t-1$, $t_{\text{evict}} \leq t - \big\lceil\frac{\lambda M}{c-\mu}\big\rceil$. Conditioned on $E$, by the properties of the RR  policy, condition in Step 16 in Algorithm \ref{algo:RR}  is satisfied for $\tau = t - \big\lceil\frac{\lambda M}{\mu-c}\big\rceil$.  It follows that in this case, the service is evicted in time-slot $t-1$ and therefore, the service is not hosted  at the edge in time-slot $t$.
	
	We thus conclude that conditioned on $F^c_{t-1} \cap E$, the service is not  hosted  at the edge during time-slot $t$. In addition, conditioned on $F^c_{t}$, the service is not fetched in time-slot $t$. 
	We now compute the expected cost incurred by the RR policy. By definition, 
	\begin{align*}
	\mathbb{E}[C_t^{\text{RR}}] =  \mathbb{E}[C_t^{\text{RR}}|G] \mathbb{P}(G) + \mathbb{E}[C_t^{\text{RR}}|G^c] \times \mathbb{P}(G^c).
	\end{align*}
	Note that, 
	$
	\mathbb{E}[C_t^{\text{RR}}|G] = \nu, \  \mathbb{E}[C_t^{\text{RR}}|G^c] \leq c+ \nu + M.
	$
	Therefore,
	\begin{align}
	\mathbb{E}&[C_t^{\text{RR}}] = \nu + (c+M) \mathbb{P}(G^c) \nonumber\\
	&\leq  \nu + \exp\left(-2\frac{(\lambda-1)^2(c-\mu)M}{\lambda  (\kappa+c_{\text{max}}-c_{\text{min}})^2}\right)+ \nonumber\\
	&(c+M) \times \bigg\lceil\frac{2\lambda M}{c-\mu}\bigg\rceil \frac{\exp\left(-2\frac{(c-\mu)^2 \frac{M}{\kappa-c}}{ (\kappa+c_{\text{max}}-c_{\text{min}})^2}\right)}{1-\exp\left(-2\frac{(c-\mu)^2}{ (\kappa+c_{\text{max}}-c_{\text{min}})^2}\right)}. 
 \label{eq:finalBound2}
	\end{align}
	We optimize over $\lambda>1$ to get the tightest possible bound. By Lemma \ref{lemma:optimal_causal} and \eqref{eq:finalBound2}, we have the result for RR.  
\end{IEEEproof}

\begin{IEEEproof}[Proof of Theorem \ref{thm:RR_stochastic_theorem}]
	Follows by Lemma \ref{lemma:optimal_causal}, \eqref{eq:finalBound}, \eqref{eq:finalBound2}, the definition of $\sigma^{\mathcal{P}}$ in \eqref{eq:efficiencyRatio}, and the fact that the cost incurred in a time-slot by any policy is upper bounded by $M+c+\nu$.
\end{IEEEproof}

\subsection{Proof of Theorem \ref{thm:fixed_TTL}}
\begin{IEEEproof}
	Throughout this proof we consider  a renting cost sequence where the cost of renting is $c_{\text{max}}$ in each time-slot.
	
	Consider the case where $\kappa < M+c_{\text{max}}$. For this setting we construct a request sequence where a request arrives in the first time-slot and no requests arrive thereafter. OPT-OFF does not fetch the service and the total cost of service per request for this request sequence under OPT-OFF is one unit. Let $C^{\text{TTL}}$ be the cost of service per request for this request sequence under  TTL. TTL fetches the service on a request arrival and stores it on local edge server for $L$ time slots. Thus the cost of service incurred by TTL per request is $C^{\text{TTL}}=1+M+Lc_{\text{max}}.$ Therefore, $\rho^{\text{ TTL}}\geq 1+M+Lc_{\text{max}}.$	
	
	Next, we consider the case where $\kappa \geq M+c_{\text{max}}$ and $Lc_{\text{max}} > M$. For this setting we construct a request sequence where $\kappa$ requests arrive in the first time-slot and no requests arrive thereafter. In this case, $C^{\text{TTL}} = \kappa + Lc_{\text{max}} + M$. Consider an alternative policy which fetches the service before the first time-slot and hosts  it for one time-slot. The total cost of service for this policy is $M+c_{\text{max}}$. It follows that $$\rho^{\text{TTL}}\geq \dfrac{\kappa+M+Lc_{\text{max}}}{M+c_{\text{max}}}.$$
	Next, we consider the case where $\kappa \geq M+c_{\text{max}}$ and $Lc_{\text{max}} \leq M$. For this setting we construct a request sequence where $\kappa$ requests arrive in time-slots $1, L+2, 2L+3, \cdots, (u-1)L+u$ and no requests arrive in the remaining time-slots. In this case, $C^{\text{ TTL}} = u(\kappa + Lc_{\text{max}} + M)$. Consider an alternative policy which fetches the service before the first time-slot and hosts  it till time-slot $(u-1)L+u$. The total cost of service for this policy is $M+((u-1)L+u)c_{\text{max}}$. It follows that $$\rho^{\text{TTL}}\geq \sup_{u \geq 1} \dfrac{u(\kappa + Lc_{\text{max}} + M)}{M+((u-1)L+u)c_{\text{max}}} = \dfrac{\kappa + Lc_{\text{max}} + M}{Lc_{\text{max}}+c_{\text{max}}}.$$
	\noindent The result follows from the three cases. 
\end{IEEEproof}
\subsection{Proof of Theorem \ref{thm:TTL_stochastic_theorem}}
\begin{IEEEproof}[Proof of Lemma \ref{lemma:TTLstochastic}]
	We compute the expectation of the cost incurred by the TTL policy in time-slot $t$. Let $G$ be the event that there are no arrivals in time-slots $\max\{t-L,1\}$ to $t-1$. Under Assumption \ref{assum_stochastic}, 
	$\mathbb{P}(G) = p_0^{\min\{L,t-1\}}.$
	For the TTL policy, conditioned on $G$, the service is not hosted  at the beginning of time-slot $t$ and all requests received in time-slot $t$ are forwarded to the back-end server. In addition, the service is fetched if at least one request is received in time-slot $t$. It follows that $\mathbb{E}[C_t^{\text{TTL}}]  = M(1-p_0) + \nu.$\\
	If the service is hosted at the edge in time-slot $t$, the TTL policy pays a rent cost of $R_t$ and up to $\kappa$ results are served at the edge. The remaining requests are forwarded to the back-end server. It follows that
	\begin{align*}
	\mathbb{E}[C_t^{\text{TTL}}|G^c]  &= \mathbb{E}[R_t] + \mathbb{E}[X_t-\min\{X_t,\kappa\}]= c + \nu- \mu.
	\end{align*}
	Note that $\mathbb{E}[C_t^{\text{TTL}}] = \mathbb{E}[C_t^{\text{TTL}}|G] \mathbb{P}(G) + \mathbb{E}[C_t^{\text{TTL}}|G^c] \mathbb{P}(G^c)=(M(1-p_0)+\nu) p_0^{\min\{L,t-1\}} + (c + \nu-\mu) (1-p_0^{\min\{L,t-1\}}). $

	Moreover, $\mathbb{E}[C_t^{\text{OPT-OFF}}] \leq \nu.$
	It follows that $\Delta^{\text{TTL}} \geq M(1-p_0)p_0^{\min\{L,t-1\}} + (c-\mu)(1-p_0^{\min\{L,t-1\}}),$
	thus proving the result.
\end{IEEEproof}
\begin{IEEEproof}[Proof of Theorem \ref{thm:TTL_stochastic_theorem}]
	Follows by Lemma \ref{lemma:TTLstochastic} and the fact that the cost per time-slot incurred by the optimal online policy is at most $\nu$.
\end{IEEEproof}
\section{Conclusions and Future Work}
In this work, we focus on designing online strategies for service hosting on edge computing platforms. We show that the widely used and studied TTL policies do not perform well in this setting. This is because, on a miss, TTL fetches the data and code needed to run the service and hosts it on the edge server, whereas, for low request arrival rates and/or high fetch cost, it is more efficient to forward all requests to the back-end server. 

In addition, we propose an online policy named RR.  Via analysis for adversarial  and  stochastic settings and simulations for synthetic and trace-based arrivals, rent costs, we show that RR  performs well for a wide array of request arrival processes  and  rent cost sequences. 




\newpage
\appendix
\section{Efficient RetroRenting}
We now discuss Efficient RetroRenting (E-RR) (Algorithm \ref{algo:RRE}), an efficient implementation of RR. 
Let $\underbar{\text{$x$}}_t$ = $\min \{x_t, \kappa\}$.
We maintain a quantity $\Delta(t)$, defined as follows:
\begin{align}\label{eq:recursive}
\Delta(t)=\min\Big\{M,\max\big\{0,\Delta(t-1)+\underbar{\text{$x$}}_t-c_t\big\}\Big\},
\end{align}
for $t\geq 1$ and $\Delta(0)=0$. Note that $\Delta(t)\in [0, M].$

\begin{algorithm}
	\caption{Efficient RetroRenting (E-RR)}\label{algo:RRE}
	\SetAlgoLined
	Input: Fetch cost $M$ units, maximum number of our service requests served by edge server($\kappa$),
	renting cost: $c_t$, number of requests: $x_t$, $t > 0$\\
	Output:  Service hosting strategy $r_{t+1}$, $t > 0$\\
	Initialize:  Service hosting variable $r_1 = 0$, $\Delta(0)=0$\\
	\For {\textbf{each} time-slot $t$}{
		$\Delta(t)=\min\Big\{M,\max\big\{0, \Delta(t-1)+\underbar{\text{$x$}}_t-c_t\big\}\Big\}$\\
		\uIf{$\Delta(t)=M$}{
			
			$r_{t+1}=1$\\
		}
		\uElseIf{$\Delta(t)=0$}{ 
			$r_{t+1}=0$\\
		}						
		\Else
		{
			$r_{t+1}=r_t$
		}				
	}
\end{algorithm}
In our next result, we show that Algorithms \ref{algo:RR} and \ref{algo:RRE} are equivalent.
\begin{lemma}\label{lem:RR_RRE}
	Algorithms \ref{algo:RR} and \ref{algo:RRE} are equivalent.
\end{lemma}
\begin{IEEEproof}
	Let $r_t^{\text{RR}}$  and $r_t^{\text{E-RR}}$ denote the renting variables in time-slot $t$ associated with Algorithms \ref{algo:RR} and \ref{algo:RRE} respectively.
	We have $r_1^{\text{RR}}=r_1^{\text{E-RR}}=0$. Let E-RR fetch the service after the end of time-slot $m$ for the first time. Therefore by Algorithm \ref{algo:RRE},
	$\Delta(t)<M$ for $1\leq t\leq m-1$ and $\Delta(m)=M$, i.e.,
	\begin{align}\label{ineq:fetch_RRE1}
	\Delta(m-1)+\underbar{\text{$x$}}_m-c_m\geq M.
	\end{align}
	Since $\Delta(t)<M$ for $1\leq t\leq m-1$, by \eqref{eq:recursive},    $\Delta(t)=\max\big\{0,  \Delta(t-1)+\underbar{\text{$x$}}_t-c_t\big\}$ for $1\leq t\leq m-1$.
	Let $i \in [1, m-1]$ be a time-slot such that $\Delta(i)=0$ and $\Delta(t)\neq 0$ for $i+1 \leq t<m$. Then, $\Delta(m-1)=\displaystyle\sum_{l=i+1}^{m-1} (\underbar{\text{$x$}}_l-c_l).$
	Using this and \eqref{ineq:fetch_RRE1} we get,
	\begin{align*}
	\displaystyle\sum_{l=i+1}^{m} \underbar{\text{$x$}}_l \geq \displaystyle\sum_{l=i+1}^{m} c_l+M.
	\end{align*}
	Thus, RR also fetches at the end of time-slot $m$.
	
	Suppose RR fetches the service after the end of time-slot $m'$ for the first time. Therefore by Algorithm \ref{algo:RR},
	$\displaystyle\sum_{l=1}^{m'} \underbar{\text{$x$}}_l \geq \displaystyle\sum_{l=1}^{m'} c_l+M$ and 
	$
	\displaystyle\sum_{l=1}^{t} \underbar{\text{$x$}}_l < \displaystyle\sum_{l=1}^{t} c_l+M.
	$
	for $1\leq t\leq m'-1$.
	Substituting $t=1,2,...m'-1$ successively in the above inequality and using $\Delta(0)=0,$ we get $\Delta(t)<M$ for $1\leq t\leq m'-1$, and
	\begin{align*}
	\Delta(m'-1)+\underbar{\text{$x$}}_m'-c_m'&\geq \displaystyle\sum_{l=1}^{m'-1} (\underbar{\text{$x$}}_l-c_l)+\underbar{\text{$x$}}_{m'}-c_{m'}\\
	&\geq\displaystyle\sum_{l=1}^{m'} (\underbar{\text{$x$}}_{m'}-c_{m'})
	\geq M
	\end{align*}
	Therefore, by \eqref{eq:recursive}, we get $\Delta(m')=M$, which is the  condition for fetching the service under E-RR at the end of time-slot $m'.$
	We thus conclude that the time-slots of first fetch by RR and E-RR are the same.
	
	After the first fetch of service by RR and E-RR at the end of time-slot $m$, let E-RR evict the service at the end of time-slot $n.$
	Therefore by Algorithm \ref{algo:RRE},
	$\Delta(t)\neq 0$ for $m+1\leq t\leq n-1$ and $\Delta(n)=0,$ i.e.,
	\begin{align}\label{ineq:evict_RRE1}
	\Delta(n-1)+\underbar{\text{$x$}}_n-c_n< 0.
	\end{align}
	Since $\Delta(t)\neq 0$ for $m+1\leq t\leq n-1$, from \eqref{eq:recursive},  we have   $\Delta(t)=\min\big\{M,  \Delta(t-1)+\underbar{\text{$x$}}_t-c_t\big\}$ for $m+1\leq t\leq n$.
	Let $j \in [m+1, n-1]$ be a time-slot such that $\Delta(j)=M$ and $\Delta(t)\neq M$ for $j+1 \leq t<n$. Then we get $\Delta(n-1)=\displaystyle\sum_{l=j+1}^{n-1} (\underbar{\text{$x$}}_l-c_l).$
	Using  the condition  $\Delta(m)=M$, we get $\Delta(n-1)= M+\displaystyle\sum_{l=j+1}^{n-1} (\underbar{\text{$x$}}_l-c_l).$
	Using this and \eqref{ineq:evict_RRE1}, we have that,
	\begin{align*}
	\displaystyle\sum_{l=j+1}^{n} \underbar{\text{$x$}}_l +M \leq \displaystyle\sum_{l=j+1}^{n} c_l.
	\end{align*}
	Thus, RR evicts at the end of time-slot $n.$
	
	Suppose RR evicts the service after the end of time-slot $n'$ for the first time after $m$. Therefore by Algorithm \ref{algo:RR},
	$\displaystyle\sum_{l=m+1}^{n'} \underbar{\text{$x$}}_l +M \leq \displaystyle\sum_{l=m+1}^{n'} c_l$ and 
	$
	\displaystyle\sum_{l=m+1}^{t} \underbar{\text{$x$}}_l +M > \displaystyle\sum_{l=m+1}^{t} c_l
	$
	for $m+1\leq t\leq n'-1$.
	Substituting $t=m+1,m+2,...n'-1$ successively in the above inequality and using $\Delta(m)=M$, we get $\Delta(t)>0$ for $m+1\leq t\leq n'-1$, i.e., 
	\begin{align*}
	\Delta(n'-1)+\underbar{\text{$x$}}_{n'}-c_{n'}&\leq \Delta(m)+ \displaystyle\sum_{l=m+1}^{n'-1} (\underbar{\text{$x$}}_l-c_l)+\underbar{\text{$x$}}_{n'}-c_{n'}\\
	&=M+\displaystyle\sum_{l=m+1}^{n'} (\underbar{\text{$x$}}_{l}-c_l)	<0.
	\end{align*}
	Therefore from \eqref{eq:recursive} we get $\Delta(n')=0$, which is the  condition for evicting the service under E-RR at the end of time-slot $n'.$
	Putting together the above results we conclude that the time-slots of first eviction by RR and E-RR are the same. The result then follows by induction.
\end{IEEEproof}

\end{document}